\documentclass[%
 a4paper,
 reprint,
superscriptaddress,
 amsmath,amssymb, amsfonts,
 aps,longbibliography, prl, %prx
]{revtex4-2}

\usepackage{graphicx}% Include figure files
\usepackage{dcolumn}% Align table columns on decimal point
\usepackage{color}
\usepackage{wrapfig}
\usepackage{bm}% bold math
\usepackage{hyperref}% add hypertext capabilities
\usepackage{lipsum}
\usepackage[caption = false]{subfig}

\newcommand{\bra}[1]{\langle #1|}
\newcommand{\ket}[1]{| #1 \rangle}
\newcommand{\op}[1]{\hat{#1}}
\newcommand{\mc}[1]{\mathcal{#1}}

\newcommand{\Tr}{\mathrm{Tr}}

\begin{document}

\preprint{}

\title{Role of Quantum Coherence in Kinetic Uncertainty Relations}

\author{Kacper Prech}
\email{kacper.prech@unibas.ch}
\affiliation{Department of Physics and Swiss Nanoscience Institute, University of Basel, Klingelbergstrasse 82, 4056 Basel, Switzerland}

\author{Patrick P. Potts}
\affiliation{Department of Physics and Swiss Nanoscience Institute, University of Basel, Klingelbergstrasse 82, 4056 Basel, Switzerland}

\author{Gabriel T. Landi}
\affiliation{Department of Physics and Astronomy, University of Rochester, P.O. Box 270171
Rochester, NY, USA}

\begin{abstract}

The kinetic uncertainty relation (KUR) bounds the signal-to-noise ratio of stochastic currents in terms of the number of transitions per unit time, known as the dynamical activity.
This bound was derived in a classical context and can be violated in the quantum regime due to coherent effects.
However, the precise connection between KUR violations and quantum coherence has so far remained elusive, despite significant investigation.
% While these inequalities, which have been originally developed for classical systems, have recently been extended to the quantum regime, the role of quantum coherence remains elusive. 
In this Letter, we solve this  problem by deriving a modified bound that exactly pinpoints how, and when, coherence might lead to KUR violations. 
Our bound is sensitive to the specific kind of unraveling of the quantum master equation. 
It, therefore, allows one to compare quantum jumps and quantum diffusion, and understand, in each case, how quantum coherence affects fluctuations.
% . Our bound involves a factor that fully captures the effect of coherence on the fluctuating current. For a density matrix commuting with the Hamiltonian of the system, the factor vanishes, and we recover the standard classical KUR. 
We illustrate our result on a double quantum dot, where the electron current is monitored either by electron jump detection or with continuous diffusive charge measurement. 
% We show that our bound constrains the fluctuations tighter than the different, previously derived quantum KUR. Our findings constitute an essential step to comprehensively understand the role that coherence plays in the behavior of open quantum systems. 

\end{abstract}

\maketitle

\textit{Introduction---}Superposition is one of the key features of quantum mechanics that distinguishes it from classical physics. While the most prominent consequences of quantum coherence are entanglement and nonlocality~\cite{Streltsov_2017, Horodecki_2009, Brunner_2014}, it also has a profound effect on dynamical properties, such as the fluctuations of currents in open quantum systems~\cite{Prech_2023, Kiesslich_2006, Ptaszynski_2018}, and  thermodynamic quantities such as heat and work~\cite{Latune_2020, Scandi_2020, Tajima_2021, Francica_2019, Santos_2019, Landi_2021, rodrigues2023nonequilibrium, Kurchan_2000, Tasaki_2020, TalknerLutz, Esposito_2009, Campisi_2011, Manzano_2018, Prech2024, Yada_2022}.
Crucially, since these fluctuations depend on two-time correlations~\cite{Landi_2023}, this effect is not necessarily related to the amount of coherence present in a quantum state, but rather to the dynamical generation and consumption of coherence in a process.
However, the precise way in which this takes place remains poorly understood.
% In this paper, we address this question by deriving a novel bound that constrains fluctuations of currents in Markovian open quantum systems in a way that precisely captures the role of coherence.

In classical systems, current fluctuations are constrained by a set of bounds, discovered over the last decade, and  collectively known as thermokinetic uncertainty relations~\cite{DiTerlizzi_2019, Garrahan_2017, Gingrich_2016, Barato_2015, Horowitz_2020, Potts_2019, Macieszczak_2018, Pietzonka_2017, Horowitz_2017, Koyuk_2020, Vo_2022, Liu_2020, Hasegawa_2019, Timpanaro_2019}. 
They  provide  lower bounds on the noise-to-signal ratio $D/J^2$, where $J$ is the average current, and $D$ (called the noise, scaled variance, or diffusion coefficient) quantifies its fluctuations.
Two prominent classes of bounds are the thermodynamic uncertainty relation (TUR)~\cite{Gingrich_2016, Barato_2015} and the kinetic uncertainty relation (KUR)~\cite{DiTerlizzi_2019, Garrahan_2017}. This Letter will focus on the latter, which reads
% types of these inequalities that have been initially derived for Markovian systems in the steady state, a Kinetic Uncertainty Relation (KUR) and a Thermodynamic Uncertainty Relation (TUR)~\cite{Gingrich_2016, Barato_2015},  The KUR is given by~\cite{DiTerlizzi_2019}
\begin{equation}
    \label{eq: Classical KUR}
     \frac{D}{J^2} \geq \frac{1}{A},
\end{equation}
where $A$ is the average dynamical activity (``freneticity'')~\cite{MAES20201} and measures the average number of transitions per unit time in a stochastic system. 
The fact that the right-hand side depends on $1/A$ means that high dynamical activities are required in order to decrease fluctuations. 
The bound, therefore, has a very practical implication in establishing the minimum activity required to achieve a certain precision.
The TUR has an analogous form to the KUR, but the bound is given in terms of the average entropy production rate in place of the dynamical activity, which is a measure of irreversibility.

In the quantum domain, however, Eq.~\eqref{eq: Classical KUR} can be violated. 
Several authors have worked to pinpoint the precise mechanisms responsible for these violations~\cite{Kiesslich_2006, Prech_2023, Ptaszynski_2018, Agarwalla_2018, Kalaee_2021, Rignon_2021, Liu_2021, Gerry_2022, Gerry_2023, Janovitch_2023, Bettmann_2023, Prech_2023, Ptaszynski_2018, Agarwalla_2018, Kalaee_2021, Cangemi_2020, Menczel_2021}. While TUR violations received a considerable amount of attention, results for the KUR were first explored recently in~\cite{Prech_2023}. 
It is noteworthy that quantum effects are not always beneficial for reducing fluctuations, and there are cases where it can actually be deleterious~\cite{Rignon_2021}.

This, therefore, begs the question of when, and how, can coherence be used to improve the precision of stochastic currents?
This led several authors to derive quantum extensions of the TUR and KUR~\cite{Hasegawa_2020, Vu_2022, Guarnieri_2019, Carollo_2019, Erker_2017, Hasegawa_2021, Hasegawa_2021b, Miller_2021, Hasegawa_2022, hasegawa2023quantum}.
These bounds are very useful in providing practical constraints. And they have also helped shed light on what new ingredients come into play when we move to the quantum domain. 
Unfortunately, they do not shed much light on the precise roles of coherence.

In this Letter we derive a new bound that holds for Markovian open quantum systems, in the presence of arbitrary quantum coherent effects. 
It replaces Eq.~\eqref{eq: Classical KUR} with 
\begin{equation}
\label{eq: Our KUR}
    \frac{D}{J^2} \geq \frac{(1 + \psi)^2}{A},
\end{equation}
where $\psi \propto [\op{H}, \op{\rho}_{\rm ss}]$ [c.f.~Eq.~\eqref{eq: psi}] is directly proportional to how much the steady-state density matrix $\op{\rho}_{\rm ss}$ fails to commute with the system Hamiltonian $\op{H}$ (i.e., to the amount of energetic coherence present in the steady state).
We refer to Eq.~\eqref{eq: Our KUR} as the $\psi$-KUR. 
A nearly identical bound also holds for the diffusive unraveling, with $(1+\psi) \to (1/2+\psi)$. In diffusive measurements, instead of directly detecting each monitored transition, their outputs are combined with strong reference currents, and their deviations are observed~\cite{Carmichael, Jacobs_2014}.
For incoherent processes $\psi = 0$, and our result reduces to the classical KUR [Eq.~\eqref{eq: Classical KUR}]. 
Conversely, for coherent processes, violations of Eq.~\eqref{eq: Classical KUR} become possible when $\psi \in [-2,0]$, while outside this interval violations are strictly not allowed.
This, therefore, unambiguously pinpoints energetic coherence as the fundamental ingredient required for KUR violations. 
The inequality in Eq.~\eqref{eq: Our KUR} also uncovers the special case $\psi = -2$, in which the original KUR holds, even though the system has coherence. 
To illustrate the significance of this special point, as well as the intuition behind Eq.~\eqref{eq: Our KUR}, we carry out a detailed analysis of a double quantum dot (DQD) model~\cite{VanDerWiel}.

% $\alpha = 1$ for currents obtained by monitoring jumps in the system and $\alpha = 1/2$ for continuously measured diffusive currents, $M$ is a dynamical activity in open quantum systems, and $\psi$ is a correction to the classical KUR bound~\eqref{eq: Classical KUR} that captures the effect of coherence in the density matrix. All these quantities are defined in detail below. Importantly, if the steady state density matrix of the system commutes with the Hamiltonian, $\psi = 0$, and our result reduces to the classical KUR. We illustrate our findings on a double quantum dot~\cite{VanDerWiel}, where a charge current is monitored either by electron jump detection or with continuous diffusive charge measurement as well as a coherently driven qubit and compare with the KUR derived in Ref.~\cite{Hasegawa_2020}.

\textit{The $\psi$-KUR---}We consider an open quantum system with a density matrix $\hat{\rho}_t$ that evolves in time according to the Lindblad master equation~\cite{Breuer, Potts_2021, Plenio_1998, Schaller_2013} ($\hbar = 1$)
\begin{equation}
\label{eq: Lindblad}
    \frac{d}{dt} \hat{\rho}_t = -i[\hat{H}, \hat{\rho}_t] + \sum_k D[\hat{L}_k] \hat{\rho}_t =: \mathcal{L} \hat{\rho}_t,
\end{equation}
where $\hat{H}$ is the Hamiltonian of the system, $\hat{L}_k$ are Lindblad jump operators, and $D[\hat{O}] \hat{\rho}_t := \hat{O} \hat{\rho}_t \hat{O}^\dagger - \frac{1}{2} \{\hat{O}^\dagger \hat{O}, \hat{\rho}_t \}$. 
We assume the system has a unique steady state $\mc{L} \op{\rho}_{\rm ss} = 0$.

The derivation of Eq.~\eqref{eq: Our KUR} is done in the Appendix. 
Here we only make explicit the main quantities involved. 
Our bound concerns generic stochastic counting observables and integrated currents $N(\tau)$, whose definition depends on the unraveling in question (specified below).
The average and scaled variance (noise) of the stochastic current $I(\tau) = dN(\tau)/d\tau$ are given by
\begin{equation}
\label{eq: Current}
    J = \frac{\text{E}[N(\tau)]}{\tau},\hspace{1cm} D = \frac{\text{Var}[N(\tau)]}{\tau},
\end{equation}
where $\text{E}[\cdot]$ denotes the expectation value and $\text{Var}[\cdot]$ the variance.
% , and $N(\tau)$ is the time integrated current, which depends on the stochastic unravelling of the master equation~\eqref{eq: Lindblad}. %The fluctuations of the current are quantified by the noise:
%\begin{equation}
%\label{eq: Diffusion}
%    D = \frac{\text{Var}[N(\tau)]}{\tau},
%\end{equation}
%where $\text{Var}[\cdot]$ denotes a variance. 
The dynamical activity reads~\cite{Hasegawa_2020, Vu_2022}
\begin{equation}
\label{eq: M}
    A = \sum_k  \Tr \{ \op{L}_k \op{\rho}_{\rm ss} \op{L}_k^\dagger \}
\end{equation}
and represents the average number of jumps per unit time in the steady state.
In turn, the factor $\psi$ in Eq.~\eqref{eq: Our KUR} is given by the expression
\begin{equation}
\label{eq: psi}
    \psi = \frac{\Tr \{ \mathcal{J} \mathcal{L}^+ \mathcal{H}\hat{\rho}_{\rm ss} \}} {J},
\end{equation}
where $\mathcal{H}\hat{\rho} := -i[\op{H}, \hat{\rho}] $, $\mathcal{L}^+$ is the Drazin inverse~\cite{Mandal_2016} of $\mathcal{L}$ (see Supplemental Material~\cite{SM} for details), and $\mathcal{J}$ is the current superoperator, i.e., $J = \text{Tr} \{ \mathcal{J} \op{\rho}\}$. We note that the superoperator $\mathcal{H}$ describes the amount of steady-state energetic coherence, whereas the Drazin inverse depends on the relaxation rates, highlighting that $\psi$ depends on the dynamics.

% Achieving a given noise-to-signal ratio thus requires a different minimal dynamical activity $A$ for different measurements (resulting in different currents). %For a given noise-to-signal ratio, the necessary value of $A$ is unambiguously set by $(1+\psi)^2$, determining a possible advantage or disadvantage with respect to incoherent dynamics indicating if KUR violation is possible with the help of coherence.

% In general, the quantum KUR~\eqref{eq: Our KUR} can be adapted to any type of unravelling, but here we focus on the two most prominent ones: the quantum jump unravelling and the diffusive unravelling~\cite{Carmichael, Landi_2023}. 
Crucially, since $\psi$ depends on $\mathcal{J}$, our bound is sensitive to both the particular current  measurement and the type of unraveling. 
For the jump unraveling, the current superoperator is given by
\begin{equation}
\label{eq: Current op}
    \mathcal{J} \hat{\rho} = \sum_k \nu_k  \op{L}_k \op{\rho} \op{L}_k^\dagger,
\end{equation}
where $\nu_k$ denotes the weight with which a jump $\op{L}_k$ changes the integrated current $N(t)$~\cite{Carmichael, Landi_2023}.
Conversely, for the diffusive unraveling ~\cite{Carmichael, Landi_2023}
%\begin{equation}
%    \frac{dN_t}{dt} = \Tr \{ \mc{J}_{\rm d} \op{\rho} \} + \sum_k \nu_k \frac{dW_{k}}{dt},
%\end{equation}
%where
\begin{equation}
\label{eq: Current diff}
    \mc{J}_{\rm d} \op{\rho} = \sum_k \nu_k \left( e^{-i \phi_k} \op{L}_k \op{\rho} + \op{\rho} \op{L}_k^\dagger e^{i \phi_k} \right) ,
\end{equation}
where $\phi_k$ are arbitrary angles.
%is the diffusive current superoperator and $dW_{k}$ is a Wiener increment~\cite{Landi_2023, Carmichael}. The corresponding average reads
%\begin{equation}
%    J_{\rm d} = \Tr \{ \mc{J}_{\rm d} \op{\rho} \}.
%\end{equation}
In this case the bound Eq.~\eqref{eq: Our KUR} holds with the replacements $(1 + \psi) \to (1/2 + \psi)$, $\mathcal{J} \to \mathcal{J}_d$, and $J \to J_d$. 

Having introduced the quantum $\psi$-KUR [Eq.~\eqref{eq: Our KUR}], we compare it with a different, previously obtained bound \cite{Hasegawa_2020}:
\begin{equation}
\label{eq: Hasegawa KUR}
    \frac{D}{J^2} \geq \frac{1}{A + \chi},
\end{equation}
which also applies in the steady state for the quantum jump unraveling of the Lindblad master equation[Eq.~\eqref{eq: Lindblad}]. For the diffusive unraveling, $1$ is replaced by $1/4$ in the numerator. Here $\chi$ is a different coherence-dependent factor (see Supplemental Material~\cite{SM} for the expression). However, as opposed to $\psi$, it does not concisely capture how much $\op{\rho}$ fails to commute with $\op{H}$. Moreover, for a given system, the factor $\chi$ is the same for all unravelings and current measurements, because it does not depend on the current superoperator $\mc{J}$, forfeiting a tailoring of the noise-to-signal bound to a particular current measurement. This is in contrast to $\psi$.%Lastly, due to the form of the right hand side of Ineq.~\eqref{eq: Hasegawa KUR}, the degree (or the relative ratio) of possible KUR violation is difficult to assess, in contrast to our KUR.~\eqref{eq: Our KUR}.
%And in contrast to~\eqref{eq: Our KUR}, the coherence-dependent term in~\eqref{eq: Hasegawa KUR} does not have the form of a multiplicative factor of the right-hand side.

\begin{figure*}[t]
    \centering
    \includegraphics[width=0.98\textwidth]{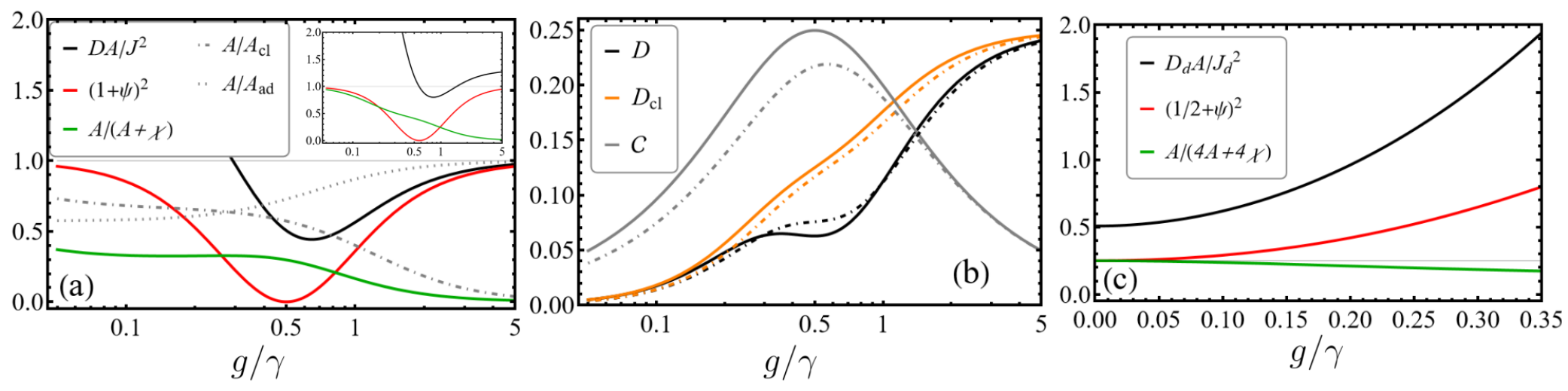}
    \caption{$\psi$-KUR in the DQD. (a) Current fluctuations in the quantum jump unraveling as a function of $g/\gamma$, with $\gamma = \gamma_L = \gamma_R$ and no dephasing ($\Gamma = 0$): (black) $DA/J^2$; (red) $(1 + \psi)^2$ [computed from Eq.~\eqref{eq: psi dqd}], which bounds the black curve according to Eq.~\eqref{eq: Our KUR}; (green) $A/(A+\chi)$ from Eq.~\eqref{eq: Hasegawa KUR}; (dashed gray) $A/A_{\rm cl}$; (dotted gray) $A/A_{\rm ad}$. Parameters: $\beta_L \mu_L = - \beta_R \mu_R = 7$ and $\epsilon = 0$. The inset shows the same plots with dephasing rate $\Gamma/\gamma = 0.3$.
    (b) The noise $D$ as a function of $g/\gamma$ for the quantum  (black) and classical (orange) models, as well as the $l_1$ norm of coherence $\mc{C}$ (gray) obtained with Eq.~\eqref{eq: norm dqd}. Solid lines correspond to the parameters of (a) (without dephasing) and dashed lines to those of the inset of (a) (with dephasing).  
    (c) Same as (a) but for the diffusive measurement of the charge difference between the quantum dots. Parameters are the same as in (a), except $\Gamma/\gamma = 1$.}
    \label{fig:panel1}
\end{figure*}

\textit{Double quantum dot---}We illustrate the $\psi$-KUR~\eqref{eq: Our KUR} on a DQD model~\cite{VanDerWiel}, where we find that $\psi\in[-2,0]$ allows for considerable violations of the classical KUR due to coherence
%(i.e. $(1+\psi)^2$ is significantly below $1$)
when the noise $D$ cannot be faithfully described by a classical model.
The system consists of left (L) and right (R) spinless quantum dots, which are weakly coupled to their respective fermionic reservoirs. The Hamiltonian is
\begin{equation}
\label{eq: H DQD}
    \op{H} = \sum_{\ell = L, R} \epsilon \op{c}_\ell^\dagger \op{c}_\ell + g \left(  \op{c}_L^\dagger \op{c}_R +  \op{c}_R^\dagger \op{c}_L \right), %+ U \op{c}^\dagger_L \op{c}_L \op{c}^\dagger_R \op{c}_R,
\end{equation}
where $\op{c}_\ell$ ($\op{c}_\ell^\dagger$) are annihilation (creation) operators of an electron in  dot $\ell = L, R$, $\epsilon$ is the occupation energy of each quantum dot, and $g$ the coherent tunnel strength. The chemical potential, inverse temperature, and coupling strength to reservoir $\ell$ are denoted by $\mu_\ell$, $\beta_\ell$, and $\gamma_\ell$, respectively.

%For the noninteracting dots ($U = 0$),
The Lindblad master equation governing the time evolution of the system is given by
\begin{equation}
\label{eq: Liouvillian dqd}
    \mathcal{L} \op{\rho} = -i[\op{H}, \op{\rho}] + \sum_\ell \mc{L}_\ell \op{\rho} +  \frac{\Gamma}{2} D[\op{c}^\dagger_L \op{c}_L - \op{c}^\dagger_R \op{c}_R] \op{\rho},
\end{equation}
where the last term denotes a dephasing in the local basis, %of the left-right tunneling
with strength $\Gamma$, while
\begin{equation}
\label{eq: Baths}
    \mc{L}_\ell \op{\rho} = \gamma_\ell \left( f_\ell D[\op{c}_\ell^\dagger] +  (1-f_\ell) D[\op{c}_\ell] \right) \op{\rho}
\end{equation}
describes the coupling to reservoir $\ell$, with 
Fermi-Dirac occupation $f_\ell := (\exp{(\beta_\ell(\epsilon - \mu_\ell))} +1)^{-1} $.  %This can be implemented, for instance, by placing a quantum point contact adjacent to the quantum dot, such that monitoring of deviations of the current in the quantum point contact, which depend on the occupation of the dot, yields a continuous diffusive charge measurement~\cite{Bettmann_2023, Goan_2001, Gurvitz_1997, Landi_2023}.

For the quantum jump unraveling, we consider a net flow of electrons from the left reservoir through the system, which has the average current $J = \gamma_L \Tr \{ f_L \op{c}_L^\dagger \op{\rho} \op{c}_L - (1-f_L) \op{c}_L \op{\rho} \op{c}_L^\dagger \}$.
%which corresponds to setting $\nu = 1$ for the $\hat{c}_L^\dagger$ jump operator, $\nu = -1$ for the $\hat{c}_L$ jump operator, and $\nu_\ell = 0$ for the remaining channels.%
%For the system without dephasing ($\Gamma =0$), in Fig.~\ref{fig:panel1}(a) as a function of $g/\gamma$ (with $\gamma := \gamma_L = \gamma_R$) 
The analytical expression for $\psi$ in this case reads
\begin{equation} \label{eq: psi dqd}
    \psi = \frac{ - 2\gamma_L \gamma_R \left( \gamma_L + \gamma_R + 2\Gamma \right)}{4g^2 \left( \gamma_L + \gamma_R \right) + \gamma_L \gamma_R \left( \gamma_L + \gamma_R + 2 \Gamma \right) }<0,
\end{equation}
which ranges between $\psi =-2$ when $g\to 0$ and $\psi=0$ when $g\to \infty$. It is, thus, always in the range $(1 + \psi)^2 <1$ such that, for this model, the DQD coherence always allows for a reduced minimal activity to sustain a fixed noise-to-signal ratio.
Fig.~\ref{fig:panel1}(a) compares $DA/J^2$ with $(1+\psi)^2$ as a function of $g/\gamma$.  
The bound is found to be tighter for large $g/\gamma$. Moreover, it is tighter than Eq.~\eqref{eq: Hasegawa KUR} for the majority of values $g/\gamma$, but not always.

The steady-state coherence, quantified by the $l_1$ norm~\cite{Baumgratz_2014} in the occupation basis, is (see Supplemental Material~\cite{SM})
\begin{equation} \label{eq: norm dqd}
    \mc{C} = \frac{2g |f_L-f_R|}{\gamma_L + \gamma_R + \Gamma}|\psi|
\end{equation}
%\begin{equation} \label{eq: norm dqd}
%\begin{aligned}
%    \mc{C} &= \frac{  4g |f_L - f_R| \gamma_L \gamma_R }{4g^2 \left( %\gamma_L + \gamma_R \right) + \gamma_L \gamma_R \left( \gamma_L + \gamma_R %+ 2\Gamma \right) }
%    \\[0.2cm]
%    &=  -\frac{2g |f_L-f_R|}{\gamma_L + \gamma_R + \Gamma}\psi.
%\end{aligned}
%\end{equation}
Thus, the range of coherent tunnel strength $g$ where $\psi$ predicts a significant contribution of coherence to the KUR violation corresponds to the peak of coherence. %Here we note that the gray line is not to be numerically compared with the other plots.

Dephasing ($\Gamma$) is found to be detrimental to loosening the KUR bound, and we find $(1 + \psi)^2 \to 1$ in the strong $\Gamma$ limit consistently with our expectations for incoherent dynamics. A similar, anticipated effect of dephasing, which features in the inset of Fig.~\ref{fig:panel1}(a), is that $DA/J^2$ obeys Eq.~\eqref{eq: Classical KUR} in a much wider range of $g/\gamma$.

\textit{Breakdown of the classical description---}%The coincidence of two dips in the red and black lines is indicative of $\psi$ quantitatively describing the actual suppression of the noise-to-signal ratio.
To gain additional insights we ask whether the noise $D$ can be captured by an effective classical model, where transport is described by a Markovian rate equation
\begin{equation}
\label{eq: Rate equation}
    \frac{d}{dt} \vec{p} = W \vec{p},
\end{equation}
where $W$ is a matrix of transition rates and $\vec{p} = [p_0, p_L, p_R, p_D]$ is a vector of probabilities that the system is empty, occupied on the left, occupied on the right, or doubly occupied. The rates arising from the coupling to the environment are obtained from Eq.~\eqref{eq: Baths}; for instance, $W_{L0} = \gamma_L f_L$. 
Conversely, the coherent tunneling is replaced by the rate
\begin{equation} \label{eq: WLR}
    W_{LR} = W_{RL} = \frac{4g^2}{\gamma_L + \gamma_R + 2 \Gamma},
\end{equation}
which can be obtained from perturbation theory~\cite{Prech_2023, Mitchison_2015, Kiesslich_2006} or by imposing that the two models should have the same average current.

Fig.~\ref{fig:panel1}(b) compares the noise in the classical and quantum models.
While the classical model describes the average current in the entire range of $g/\gamma$, there is a discrepancy in the noise $D$, which appears precisely where the coherence in Eq.~\eqref{eq: norm dqd} has a peak. In this regime, the classical model fails to reproduce the reduction of the noise predicted by the quantum master equation. This coincides with the range where $(1+\psi)^2 < 1$, i.e., where the $\psi$-KUR in Eq.~\eqref{eq: Our KUR} differs from the KUR in Eq.~\eqref{eq: Classical KUR}.
%the range where $(1+\psi)^2 \simeq 0$ [Fig.~\ref{fig:panel1}(a)].
%The factor $\psi$ thus quantifies when energetic coherence plays a contributing role in overcoming the KUR.
%The classical KUR~\eqref{eq: Classical KUR} can be recovered in the limit of incoherent dynamics as a result of $[\op{H}, \op{\rho}] = 0$, implying $\psi = 0$. 
%Advantage in overcoming the KUR stemming from energetic coherence in the density matrix implies $(1+\psi)^2 < 1$. 
%This coincides with the regime where the noise cannot be described by the effective classical model.
%In contrast, when $D$ is well captured by the classical model, and therefore we should not anticipate KUR violations, $(1+\psi)^2\simeq 1$ implying that any energetic coherence in the density matrix does not lead to KUR violations. 
%We note that in contrast to our $\psi$-KUR, the KUR in Ineq.~\eqref{eq: Hasegawa KUR} becomes very loose as $g/\gamma$ becomes large.

%Away from this regime, however, the fluctuations of the current are faithfully described by the classical rate equation~\eqref{eq: Rate equation}. This effect was investigated in detail in Refs.~\cite{Prech_2023, Kiesslich_2006}. With a finite dephasing, the discrepancy between the classical and the quantum models naturally shrinks.

While the classical model reproduces the noise for both small and large $g/\gamma$, these limits are quite different in nature. For large $g$, the Lindblad jumps provide the bottleneck for transport and, thus, determine the average current and the noise. Indeed, in this regime  $p_L - p_R \simeq 0$ can be adiabatically eliminated, and the classical model reduces to a three-state model $\vec{p} = [p_0, p_L + p_R, p_D]$ (see Supplemental Material~\cite{SM} for details). In this limit, coherence is suppressed and $\psi = 0$. In the limit of small $g/\gamma$, transport is dominated by the coherent tunneling, which now provides the bottleneck, but this tunneling can be captured by the perturbative rate $W_{LR}$. In this case, we obtain $\psi\rightarrow -2$, which also constitutes a classical limit where the $\psi$-KUR reduces to Eq.~\eqref{eq: Classical KUR}. As shown in the Appendix, since the current can be expressed as a series of conductances containing both perturbative rates and Lindblad jump rates, we find $\psi \in [-2,0]$.

%More generally, as we detail in the Appendix, when transport is dominated by the Lindblad jumps in the master equation, we obtain $\psi\rightarrow 0$. When transport is dominated by a coherent process that can be described using a perturbative rate, we obtain $\psi\rightarrow -2$, which also constitutes a classical limit. Whenever the current can be expressed as a series of conductances containing both perturbative rates and jump rates, then $\psi \in [-2,0]$. This is the case for the DQD, where $J=(\gamma_L^{-1}+\gamma_R^{-1}+W_{LR}^{-1})^{-1}(f_L-f_R)$.

It is important to point out that the dynamical activity of the classical model differs from that of the quantum model given in Eq.~\eqref{eq: M}. Indeed, the classical model implies the bound $D/J^2\geq 1/A_{\rm cl}$ with the dynamical activity $A_{\rm cl}= A + (W_{LR} -\Gamma/2) (p_L + p_R)$ [see Fig.~\ref{fig:panel1}(a)]. The term proportional to $\Gamma$ is subtracted, because dephasing jumps contribute to the dynamical activity of the quantum model but not in the classical rate equation. Nonetheless, for small $\Gamma$, in the regimes where the DQD behaves classically, the quantity $A$ is the relevant  quantity that bounds the signal-to-noise ratio according to Eq.~\eqref{eq: Classical KUR}.
For small $g/\gamma$ the contribution from $W_{LR}$ is negligible, and $A_{\rm cl}\simeq A$. For large $g/\gamma$, interdot transitions are very rapid but do not influence the noise. As mentioned above, the relevant classical dynamics may be described by a coarse-grained three-state model. The dynamical activity of this model, $A_{\rm ad}$, tends to $A$ for large $g/\gamma$ [see Fig.~\ref{fig:panel1}(a)]. Therefore, in the regimes where $D$ is captured by the classical model, the $\psi$-KUR in Eq.~\eqref{eq: Our KUR} provides the relevant bound for the signal-to-noise ratio. We note that, in contrast to the $\psi$-KUR, the KUR in Eq.~\eqref{eq: Hasegawa KUR} becomes very loose as $g/\gamma$ becomes large as the denominator on the right-hand side approaches $A_{\rm cl}$. We note that Eqs.~\eqref{eq: Our KUR} and \eqref{eq: Hasegawa KUR} can be combined to obtain a tighter bound.

\textit{Diffusive charge measurement---} We consider next the diffusive measurements of the charge difference between the dots, which can be implemented using a quantum point contact~\cite{Bettmann_2023, Goan_2001, Gurvitz_1997, Landi_2023}. The resulting diffusive current is $J_{\rm d} =  \sqrt{2\Gamma} \Tr\{ (\op{c}_L^\dagger \op{c}_L - \op{c}_R^\dagger \op{c}_R) \op{\rho} \} $, where we assumed that all dephasing in Eq.~\eqref{eq:  Liouvillian dqd} is due to the measurement. %The ratio $D_{\rm d}M/J_{\rm d}^2$ (black) and $(1/2 + \psi)^2$ (red) are shown in Fig.~\ref{fig:panel1}(c) as a function of $g/\gamma$, illustrating validity of the quantum KUR~\eqref{eq: Our KUR} as well as that $\psi$, which is given by the expression
The $\psi$-KUR is illustrated in Fig.~\ref{fig:panel1}(c), where now 
\begin{equation} \label{eq: psi dqd diff}
    \psi = \frac{  8 g^2 \left( \gamma_L + \gamma_R  \right)}{4g^2 \left( \gamma_L + \gamma_R \right) + \gamma_L \gamma_R \left( \gamma_L + \gamma_R + 2 \Gamma \right) }.
\end{equation}
In this case $\psi>0$ always, resulting in a tighter bound than the KUR [i.e., no violations of Eq.~\eqref{eq: Classical KUR} are allowed]. 
One of the key features of the $\psi$-KUR is that it is unraveling dependent. 
For instance, we can contrast our result with the quantum bound in Eq.~\eqref{eq: Hasegawa KUR} (green line), which exhibits opposite behavior when compared to Eq.~\eqref{eq: Our KUR}, thus representing a looser constraint on the noise-to-signal ratio. Interestingly, diffusive charge measurement is not the only example where we find $\psi>0$. It happens also for the jump current, where we count only electrons entering the system from the left reservoir, which corresponds to the current $J = \gamma_L f_L \Tr \{ \op{c}_L^\dagger \op{\rho} \op{c}_L \}$ (see Supplemental Material~\cite{SM}).

\begin{figure}
    \centering
    \includegraphics[width=0.4\textwidth]{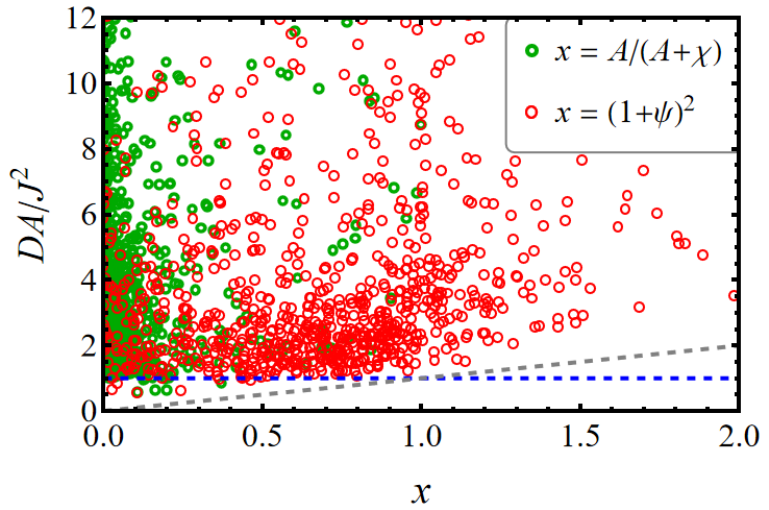}
    \caption{A scatter plot of $DA/J^2$ against $(1+\psi)^2$ (red circles) and $A/(A + \chi)$ (green circles) for 1000 random networks of a five-level quantum system. Each transition $\ket{n} \leftrightarrow \ket{k}$ is either coherently or dissipatively connected. Coherent tunneling strength $g_{nk} = g_{kn}$ and jumping rates $\gamma_{nk}$ are sampled from a uniform $U[0, 3]$, whereas $\gamma_{kn} = \gamma_{nk} e^{-\sigma_{nk}}$, with $\sigma_{nk}$ sampled from $U[3, 5]$. The current is defined along a single edge with antisymmetric weights $\pm1$. The gray dashed line represents $x$, i.e., the $\psi$-KUR bound in Eq.~\eqref{eq: Our KUR} for red circles and the bound in Eq.~\eqref{eq: Hasegawa KUR} for green circles. The blue dashed line is $1$, i.e., the classical KUR bound in Eq.~\eqref{eq: Classical KUR}.} 
    \label{fig: random network}
\end{figure}

\textit{Random network of states---} To illustrate the $\psi$-KUR in more general settings, we numerically investigate a five-level system, where each transition $\ket{n} \leftrightarrow \ket{k}$ between the computational basis states is realized either by a coherent interaction $g_{nk}  \ket{n} \bra{k} $ or jump operators $
\sqrt{\gamma_{nk}} \ket{n} \bra{k}$ and $
\sqrt{\gamma_{kn}} \ket{k} \bra{n}$. Values of $g_{nk} = g_{kn} $ and $\gamma_{nk}$ are randomly sampled from a uniform distribution $U[0, 3]$, whereas $\gamma_{kn} = \gamma_{nk} e^{-\sigma_{nk}}$, with $\sigma_{nk}$ sampled form $U[3, 5]$ to ensure a strong bias. We consider an antisymmetric current for a single transition $\ket{0} \leftrightarrow \ket{1}$, meaning that $\sqrt{\gamma_{10}} \ket{1} \bra{0}$ and $\sqrt{\gamma_{01}} \ket{0} \bra{1}$ have weights $1$ and $-1$, respectively, whereas all other jumps are not counted toward the current. 

The red circles in Fig.~\ref{fig: random network} show $DA/J^2$~vs~$(1+ \psi)^2$ for $1000$ randomly sampled networks. These points illustrate that the factor $\psi$ may be both positive and negative, and that the $\psi$-KUR provides a relevant bound also for more complex systems. The green circles show $DA/J^2$~vs.~$A/(A+\chi)$ for the same network. It is noteworthy that many green circles fall close to $x = 0$, where Eq.~\eqref{eq: Hasegawa KUR} reduces to the trivial bound $DA/J^2 \geq 0$.

\textit{Conclusions and outlook---}We derived an unraveling-dependent quantum KUR that holds for Markovian open quantum systems in the steady state, and includes a factor ($\psi$) that captures the effect of energetic coherence in the density matrix. This bound pinpoints precisely how coherence may or may not allow for violations of the classical KUR. The physical significance of our results is substantiated by illustrating the KUR and the meaning of $\psi$ on the DQD with with two different unravelings corresponding to different measurements. 
Interesting future questions include deriving thermokinetic uncertainty relations under strong system-environment couplings or for a unitary description of system and environment, investigating the transient regime~\cite{blasi2024, Vu_2022, bourgeois2024finitetime}, extending the results to the first passage times~\cite{Vu_2022, Kewming_2023}, exploring the implications of our findings on the precision of clocks~\cite{Meier_2023, Culhane_2024, Erker_2017}, and generalizing a newly established clock uncertainty relation~\cite{prech2024optimaltimeestimationclock, macieszczak2024ultimatekineticuncertaintyrelation}, which constitutes a generally tighter bound than the classical KUR with the average residual time in place of the dynamical activity, to the quantum regime. Since there is a growing interest in current fluctuations in non-Markovian settings~\cite{Brenes_2023, bettmann2024quantumstochasticthermodynamicsmesoscopicleads}, with a potential development of a theory of unravelings in non-Markovian master equations~\cite{Li}, our methods could be useful to find similar bounds on the signal-to-noise ratio.

\begin{acknowledgments}
\textit{Acknowledgements---}P.P.P. and K.P. acknowledge funding from the Swiss National Science Foundation (Eccellenza Professorial Fellowship PCEFP2\_194268). 
\end{acknowledgments}

\appendix

\section{Appendix: Derivation of $\psi$-KUR [Eq.~\eqref{eq: Our KUR}]}

To derive the $\psi$-KUR given in Eq.~\eqref{eq: Our KUR}, we consider the following deformation of the jump operators in the Lindblad master equation [Eq.~\eqref{eq: Lindblad}]: $\op{L}_k \to \op{L}_{k, \theta} :=  \sqrt{1+\theta} \op{L}_k$. This results in the modified Lindblad master equation
\begin{equation}
\label{eq: Lindblad theta}
    \frac{d}{dt} \hat{\rho}_t = -i[\hat{H}, \hat{\rho}_t] + (1 + \theta) \sum_k D[\hat{L}_{k}] \hat{\rho} =: \mathcal{L}_\theta \hat{\rho}_t,
\end{equation}
such that we recover the original master equation when $\theta \to 0$. 
Contrary to the method used to derive Eq.~\eqref{eq: Hasegawa KUR}, here the deformation is only in the jump operators. It, therefore, does not amount to a homogeneous scaling of time. 
We use a generalized quantum Cramér-Rao bound~\cite{Helstrom, Hotta_2004}
\begin{equation}
\label{eq: Cramer Rao}
    {\rm Var}_\theta[N(\tau)]_{\theta = 0} \geqslant \frac{\big(\partial_\theta {\rm E}_\theta[N(\tau)]_{\theta = 0})^2}{\mc{I}(\theta \to 0)},
\end{equation}
where ${\rm E}_\theta[N(\tau)]$ and ${\rm Var}_\theta[N(\tau)]$ denote the expectation value and the variance of $N(\tau)$ corresponding to the distorted dynamics [Eq.~\eqref{eq: Lindblad theta}], and $\mc{I}(\theta)$ is the quantum Fisher information (QFI) of the parameter $\theta$. In Eq.~\eqref{eq: Cramer Rao}, $N(\tau)$ plays the role of an estimator for the parameter $\theta$. While it may not be a good estimator (it is generally biased), it nevertheless is a possible estimator and, thus, obeys the Cramér-Rao bound.
While Eq.~\eqref{eq: Cramer Rao} holds universally, analytically computing all expressions therein is often not straightforward, as evidenced by this Letter. Similarly, finding measurements that saturate it is highly nontrivial.

Using the formalism of Ref.~\cite{Gammelmark_2014} to compute the QFI of continuously monitored open quantum systems, we find
\begin{equation}
\label{eq: QFI}
    \mc{I}(\theta \to 0) = \tau A.
\end{equation}
The expectation value of the time-integrated current is given by
\begin{equation}
\label{eq: Current theta}
    {\rm E}_\theta[N(\tau)] =
    \begin{cases}
        & (1+\theta) \int_0^\tau dt \text{Tr} \{ \mathcal{J} e^{\mathcal{L}_\theta \tau} \op{\rho}   \} \text{   (jump current)}\\[0.3cm]
        & \sqrt{1+\theta} \int_0^\tau dt \text{Tr} \{ \mathcal{J} e^{\mathcal{L}_\theta \tau} \op{\rho}   \} \text{   (diffusive current)}.
    \end{cases} 
\end{equation}
Due to a different prefactor, we obtain two expressions for its partial derivative with respect to $\theta$:
\begin{equation}
\label{eq: Derivative Current theta}
    \partial_\theta {\rm E}_\theta\big[N(\tau)\big]_{\theta = 0} =
    \begin{cases}
        &J \tau(1 + \psi) \text{   (jump current)}\\[0.2cm]
        &J \tau ( \frac{1}{2} + \psi) \text{   (diffusive current)}
    \end{cases}.
\end{equation}
Using the relation ${\rm Var}_\theta[N(\tau)]_{\theta = 0} =  D \tau$ and inserting Eqs.~\eqref{eq: QFI} and~\eqref{eq: Derivative Current theta} into the quantum Cramér-Rao bound [Eq.~\eqref{eq: Cramer Rao}] leads to our $\psi$-KUR [Eq.~\eqref{eq: Our KUR}]. See Supplemental Material~\cite{SM} for a derivation of Eqs.~\eqref{eq: QFI} and~\eqref{eq: Derivative Current theta}.

The classical KUR [Eq.~\eqref{eq: Classical KUR}] can be recovered in the limit of incoherent dynamics as a result of $[\op{H}, \op{\rho}] = 0$, implying $\psi = 0$. 
For the DQD, this is what happens in the limit $g \to \infty$. As discussed in the main text, we may also recover $(1 + \psi)^2 = 1$ when $\psi = -2$, which happens for the DQD in the limit $g\to 0$ where both $[\op{H}, \op{\rho}]$ as well as $J$ tend to zero, c.f. Eq.~\eqref{eq: psi dqd}. This observation may be understood by considering the rates in the classical model. In the limit of large $g$, the couplings to the bath become the bottleneck that dominates transport, and the current reduces to $J=\gamma_L\gamma_R(f_L-f_R)/(\gamma_L+\gamma_R)$. Under the rescaling in Eq.~\eqref{eq: Lindblad theta}, $\text{E}_\theta[N(\tau)] = J\tau (1+\theta)$, which results in $\psi=0$ from Eq.~\eqref{eq: Derivative Current theta}. In contrast, in the limit of small $g$, interdot tunneling with rate $ W_{LR}$ provides the bottleneck and the current reduces to $J=W_{LR}(f_L-f_R)$. Since $W_{LR}$ is inversely proportional to the Lindblad jump rates, the rescaled integrated current reduces to $\text{E}_\theta[N(\tau)] = J\tau/(1+\theta)\simeq J\tau(1-\theta)$ for small $\theta$, which results in $\psi=-2$. For arbitrary interdot couplings, the current can be written as a series of conductances
\begin{equation}
    J=(\gamma_L^{-1}+\gamma_R^{-1}+W_{LR}^{-1})^{-1}(f_L-f_R).
\end{equation}
The rescaled current then reads, for small $\theta$, 
\begin{equation}
    \text{E}_\theta[N(\tau)] =J\tau \left(1+\theta\frac{W_{LR}(\gamma_L+\gamma_R)-\gamma_L\gamma_R}{W_{LR}(\gamma_L+\gamma_R)+\gamma_L\gamma_R}\right),
\end{equation}
which imposes in $-2\leq\psi\leq 0$. As we detail in Supplemental Material~\cite{SM}, the same restrictions for $\psi$ hold whenever the current can be written as a series of conductances including both perturbative rates as well as Lindblad jump rates.

It is interesting to examine Eqs.~\eqref{eq: Our KUR} and~\eqref{eq: Hasegawa KUR} as a quantum Cramér-Rao bound, c.f. Eq.~\eqref{eq: Cramer Rao}. In Eq.~\eqref{eq: Our KUR}, the quantum correction $\psi$ arises from the numerator, i.e., the bias of the estimator. The bound becomes trivial (i.e., the right-hand side vanishes) when the estimator does not depend on $\theta$, which happens for $\psi=-1$ [c.f.~Eq.~\eqref{eq: Derivative Current theta}]. In this regime, where coherence plays a strong role, we can, thus, not expect our bound to be tight, as the fluctuations generally remain finite even when they do not contain information on $\theta$. In contrast, the quantum correction $\chi$ in Eq.~\eqref{eq: Hasegawa KUR} arises from the quantum Fisher information, i.e., the denominator in Eq.~\eqref{eq: Cramer Rao} \cite{Hasegawa_2020}. The bound becomes loose when there is a large amount of information on $\theta$ in the output of the system. This generally happens when the system hosts fast processes, which allow for estimating time precisely, since in Ref.~\cite{Hasegawa_2020}, $\theta$ corresponds to a rescaling of time. This explains why Eq.~\eqref{eq: Hasegawa KUR} becomes loose for large $g$ in Fig.~\ref{fig:panel1}.

% More generally, when transport is dominated by the Lindblad jumps in the master equation, we obtain $\psi\rightarrow 0$. When transport is dominated by a coherent process that can be described using a perturbative rate, we obtain $\psi\rightarrow -2$, which also constitutes a classical limit. Whenever the current can be expressed as a series of conductances containing both perturbative rates and jump rates, then $\psi \in [-2,0]$. This is the case for the DQD, where $J=(\gamma_L^{-1}+\gamma_R^{-1}+W_{LR}^{-1})^{-1}(f_L-f_R)$.

%The term $1 + \psi$ originates from $\partial_\theta {\rm E}_\theta\big[N(\tau)\big]_{\theta = 0} / {\rm E}\big[N(\tau)\big] $, and the bottleneck of the current is set by the smallest classical rate (here $\gamma_{L, R}$) or quantum strength (here $g$). Specifically, the current is primarily proportional either to $\gamma_{L, R}$ or $ W_{LR} = 4g^2/(\gamma_L + \gamma_R + 2\Gamma)$, cf. Eq.~\eqref{eq: WLR}. Since in the parameter imprinting~\eqref{eq: Lindblad theta} we leave the Hamiltonian unchanged, $\gamma_{L, R} \to \gamma_{L, R}$ and $W_{LR} \to W_{LR}/(1 + \theta) \approx W_{LR}(1-\theta)$ for small $\theta$. Therefore, when $g$ is the smallest rate, $\partial_\theta {\rm E}_\theta\big[N(\tau)\big]_{\theta = 0} \to - {\rm E}\big[N(\tau)\big]$, leading to $\psi \to -2$. Whereas when $g $ is the largest rate, $\partial_\theta {\rm E}_\theta\big[N(\tau)\big]_{\theta = 0} \to  {\rm E}\big[N(\tau)\big]$, resulting in $\psi \to 0$, in agreement with Eq.~\eqref{eq: psi dqd} approaching $0$ when $g$ dominates over $\gamma_{L, R}$ and $\Gamma$.

\bibliography{refs}

%\end{document}

\widetext
\pagebreak
\begin{center}
\textbf{\large Supplemental Material: Role of Quantum Coherence in Kinetic Uncertainty Relations}
\end{center}
\begin{center}
Kacper Prech$^{1,*}$, Patrick P. Potts$^1$, and Gabriel T. Landi$^{2}$ 
\end{center}
\begin{center}
$^1$\textit{Department of Physics and Swiss Nanoscience Institute, University of Basel, Klingelbergstrasse 82, 4056 Basel, Switzerland}
\flushleft
\hspace{1.9cm}$^2$\textit{Department of Physics and Astronomy, University of Rochester, P.O. Box 270171
Rochester, NY, USA}
\flushleft
\hspace{1.9cm}$^*$ kacper.prech@unibas.ch
\end{center}

%%%%%%%%%% Merge with supplemental materials %%%%%%%%%%
%%%%%%%%%% Prefix a "S" to all equations, figures, tables and reset the counter %%%%%%%%%%
\setcounter{equation}{0}
\setcounter{figure}{0}
\setcounter{table}{0}
\setcounter{page}{1}
\makeatletter
\renewcommand{\theequation}{S\arabic{equation}}
\renewcommand{\thefigure}{S\arabic{figure}}
\renewcommand{\thetable}{S\arabic{table}}
%\renewcommand{\bibnumfmt}[1]{[S#1]}
%\renewcommand{\citenumfont}[1]{S#1}
%%%%%%%%%% Prefix a "S" to all equations, figures, tables and reset the counter %%%%%%%%%%

In Sec. I we present the derivation of the quantum KUR, Eq.~\eqref{eq: Our KUR} in the main text. In Sec. II we discuss a restriction on $\psi$ when the corresponding current can be expressed as a series of conductances. In Sec. III we state the KUR of Ref.~\cite{Hasegawa_2020}, Eq.~\eqref{eq: Hasegawa KUR} in the main text. In Sec. IV we provide details about the Drazin inverse and how to compute it, provide useful relations about vectorization of quantum operators, and explain how to compute the noise $D$. Details of the double quantum dot are presented in Sec. V. The classical model, including the rate equation and the adiabatic elimination for large $g$, are discussed in Sec. VI. In Sec. VII we give details about the example with the random network of states. Lastly, in Sec. VIII, we introduce and investigate a coherently driven qubit, an additional, further example illustrating our results.

\section{I. Derivation of the quantum KUR}  \label{supp: Derivation}

In this section, we present a derivation of Eqs.~\eqref{eq: QFI} and~\eqref{eq: Derivative Current theta} in the main text.

\subsection{Calculation of the QFI $\mc{I}(\theta)$}
In order to derive Eq.~\eqref{eq: QFI} in the main text, the QFI is computed using the result of Gammelmark and M{\o}lmer~\cite{Gammelmark_2014}:
\begin{equation}\label{GMo_QFI}
\begin{split}
    \mc{I}(\theta) =& 4 \tau \sum_k \Tr\{(\partial_\theta \op{L}_{k,\theta})^\dagger (\partial_\theta \op{L}_{k,\theta}) \op{\rho}_{\rm ss} \}
    - 4 \tau \left(  \Tr \{  \mathcal{L}_L \mathcal{L}^+ \mathcal{L}_R \op{\rho}_{\rm ss} \} + \Tr \{ \mathcal{L}_R \mathcal{L}^+ \mathcal{L}_L \op{\rho}_{\rm ss} \} \right),
\end{split}
\end{equation}
where 
\begin{equation}
\begin{split}
    \mathcal{L}_L \op{\rho} &= -i \partial_\theta \op{H} \op{\rho} - \frac{1}{2} \sum \limits_k \partial_\theta (\op{L}_{k,\theta}^\dagger \op{L}_{k,\theta}) \rho + \sum_k (\partial_\theta \op{L}_{k,\theta}) \op{\rho} \op{L}_{k,\theta}^\dagger, \\
    \mathcal{L}_R \op{\rho} &= i \op{\rho} \partial_\theta \op{H}- \frac{1}{2} \sum\limits_k \op{\rho} \partial_\theta (\op{L}_{k,\theta}^\dagger \op{L}_{k,\theta})  + \sum_k \op{L}_{k,\theta} \op{\rho} (\partial_\theta \op{L}_{k,\theta})^\dagger.
\end{split}
\end{equation}
In the parametrization considered in Eq.~\eqref{eq: Lindblad theta} in the main text, only jump operators, $L_{k,\theta} = \sqrt{1+\theta} L_k$, are dependent on $\theta$. Therefore, in the limit $\theta \to 0$, the superoperators $\mathcal{L}_{L,R}$ reduce to 
\begin{equation}
\begin{split}
    \mathcal{L}_L \op{\rho} =& \frac{1}{2} \sum_k \Big( \op{L}_k \op{\rho} \op{L}_k^\dagger - \op{L}_k^\dagger \op{L}_k  \op{\rho} \Big), \\
    \mathcal{L}_R \op{\rho} =& \frac{1}{2} \sum_k \Big( \op{L}_k \op{\rho} \op{L}_k^\dagger - \op{\rho} \op{L}_k^\dagger \op{L}_k \Big).
\end{split}
\end{equation}
One can observe, however, that these operators are traceless, $\Tr \{ \mathcal{L}_{L,R} \op{\rho} \}= 0$.
Therefore, the last two terms in Eq.~\eqref{GMo_QFI} vanish, and in the limit $\theta \to 0$, the QFI reduces to
\begin{equation}\label{SM_QFI_final}
    \mc{I}(\theta \to 0) = \tau \sum_k \Tr \{ L_k^\dagger L_k \op{\rho}_{\rm ss} \} = \tau A, 
\end{equation}
where $A$ is the dynamical activity, see Eq.~\eqref{eq: M} in the main text.

\subsection{Calculation of $\partial_\theta {\rm E}_\theta [N(\tau)] $}
Here we show how to obtain Eq.~\eqref{eq: Derivative Current theta} in the main text, starting from the jump current. For $\theta = 0$, the average of $N(\tau)$ is given by
\begin{equation}
\begin{split}
    {\rm E} [N(\tau)] =& \int_0^\tau \text{E}[dN_t] = \sum_k \int_0^\tau dt~\nu_k \Tr \{ \op{L}_k \op{\rho}_t \op{L}_k^\dagger \} \\
    =& \sum_k \int_0^\tau dt~\nu_k \Tr \{ \op{L}_k^\dagger \op{L}_k e^{\mathcal{L}t } \op{\rho}_0 \},
\end{split}
\end{equation}
where the density matrix explicitly includes a time dependence.  
The corresponding quantity in the deformed dynamics is given by
\begin{equation} \label{eq: expectation theta}
    \text{E}_\theta[N(\tau)] = (1+\theta) \sum_k \int_0^\tau dt~\nu_k \Tr \{ \op{L}_k^\dagger \op{L}_k e^{\mathcal{L}_\theta t } \op{\rho}_0 \},
\end{equation}
with $\op{\rho}_0$ being the initial state.
Since $\theta$ appears both in the prefactor and in the $e^{\mathcal{L}_\theta t}$ term, we find
\begin{equation}
\label{eq: Terms current}
\begin{split}
    \partial_\theta \text{E}_\theta [N(\tau)] =& \sum_k \int_0^\tau dt~\nu_k \Tr \{ \op{L}_k^\dagger \op{L}_k e^{\mathcal{L}_\theta t } \op{\rho}_0 \} \\
    &+ (1+\theta) 
    \sum_k \int_0^\tau dt~\nu_k \Tr \{ \op{L}_k^\dagger \op{L}_k \big(\partial_\theta e^{\mathcal{L}_\theta t }\big) \op{\rho}_0 \}.
\end{split}
\end{equation}
Since this expression is written in terms of the time evolution from the initial state, there is no $\theta$ dependence in $\op{\rho}_0$.
Recall that we are only interested in the limit $\theta\to 0$. Thus, for instance, the first term, taking $\op{\rho}_0$ to be the steady state, simplifies to
\begin{equation}
    \lim_{\theta \to 0} \sum_k \int_0^\tau dt~\nu_k \Tr \{ \op{L}_k^\dagger \op{L}_k e^{\mathcal{L}_\theta t } \op{\rho}_0 \} = \tau J, 
\end{equation}
where $J$ is the average current, see Eq.~\eqref{eq: Current} in the main text.
To simplify the second term in Eq.~\eqref{eq: Terms current}, we apply a Dyson series expansion to $\mathcal{L}_\theta = \mathcal{L} + \theta \mathcal{D}$, which leads to 
\begin{equation}
    e^{\mathcal{L}_\theta t} = e^{\mathcal{L}t} + \theta \int_0^\tau dt_1 e^{\mathcal{L} (t-t_1)} \mathcal{D} e^{\mathcal{L} t_1} + \mathcal{O}(\theta^2).  
\end{equation}
Thus, to leading order, we have 
\begin{equation}
    \partial_\theta e^{\mathcal{L}_\theta t} = \int_0^t dt_1~e^{\mathcal{L}(t-t_1)} \mathcal{D} e^{\mathcal{L} t_1} + \mathcal{O}(\theta). 
\end{equation}
Inserting this expression into Eq.~\eqref{eq: Terms current} results in  
\begin{equation}
     \partial_\theta \text{E}_\theta [N(\tau)]_{\theta = 0} = \tau J + \sum_k \nu_k \int_0^\tau dt \int_0^t dt_1 ~\Tr \{ \op{L}_k^\dagger \op{L}_k e^{\mathcal{L}(t-t_1)} \mathcal{D} e^{\mathcal{L} t_1} \op{\rho}_0 \}.
\end{equation}
Since $\op{\rho}_0$ is the steady state, we have $e^{\mathcal{L} t_1} \op{\rho}_0 = \op{\rho}_0$, leading to  
\begin{equation}\label{SM_deriv_partial1}
 \partial_\theta \text{E}_\theta [N(\tau)]_{\theta = 0} = \tau J  + \sum_k \nu_k \int_0^\tau dt \int_0^t dt_1 ~\Tr \{ \op{L}_k^\dagger \op{L}_k e^{\mathcal{L}(t-t_1)} \mathcal{D} \op{\rho}_0 \}.
\end{equation}
To evaluate the second term on the right-hand-side of the equation above, we switch to a vectorized notation. The trace and the steady state are the left and right eigenvectors of the Liouvillian's zero eigenvalue:
\begin{equation}\label{SM_vectorized_steady_state}
    \langle 1 | \mathcal{L} = 0 
    \quad \text{and} \quad 
    \mathcal{L} |\rho\rangle = 0,
\end{equation}
respectively.
We  assume  the steady state is unique and the Liouvillian is diagonalizable. The latter is not necessary, but does simplify what follows.
The remaining eigenvalue and eigenvector pairs will be denoted as
\begin{equation}
\mathcal{L}|x_j\rangle = \lambda_j |x_j\rangle 
    \quad \text{and} \quad 
    \langle y_j | \mathcal{L} = \lambda_j \langle y_j |.
\end{equation}
It is also assumed that the system is stable, which implies that ${\rm Re}(\lambda_j)<0$.
Moreover, orthogonality implies that the eigenvectors should satisfy 
\begin{equation}\label{eigenvectors}
    \langle 1 | \rho \rangle = 1, 
    \quad \langle y_j | x_k \rangle = \delta_{j,k}, 
    \quad \text{and} \quad \langle 1 | x_j \rangle = \langle y_j |\rho\rangle = 0.
\end{equation}
This allows us to write the following operator decompositions:
\begin{equation}
\label{Eye_eigen}
I = |\rho\rangle\langle 1 | + \sum\limits_j |x_j\rangle\langle y_j |,
\end{equation}
\begin{equation}
    \label{L_eigen}
\mathcal{L} = \sum\limits_j \lambda_j |x_j\rangle\langle y_j|, 
\end{equation}
and
\begin{equation}
\label{expL_eigen}
e^{\mathcal{L}t} = |\rho\rangle\langle 1 | + \sum\limits_j e^{\lambda_j t} |x_j\rangle\langle y_j |,    
\end{equation}
where $I$ in Eq.~\eqref{Eye_eigen} is the identity matrix in the vectorized space. 
With these definitions, the double integral in Eq.~\eqref{SM_deriv_partial1} can now be written as 
\begin{equation}
    \int_0^\tau dt \int_0^t dt_1~e^{\mathcal{L}(t-t_1)} =  \sum\limits_j \left( \frac{e^{\lambda_j \tau} - \lambda_j \tau - 1}{\lambda_j^2}\right) |x_j\rangle\langle y_j|  
     + \frac{\tau^2}{2} |\rho\rangle\langle 1| .
\end{equation}
This result holds for arbitrary $\tau$ and could be useful, for instance, if one wishes to extend the present calculations beyond the asymptotic regime.
In our case, however, we are only interested in large $\tau$, or, more precisely, in $\lambda_j \tau \gg 1$ for all $\lambda_j$. 
As a consequence, the integral reduces to
\begin{equation}
\label{eq: intergral final}
    \int_0^\tau dt \int_0^t dt_1~e^{\mathcal{L}(t-t_1)} = \frac{\tau^2}{2} |\rho\rangle\langle 1| - \tau~\mathcal{L}^+,
\end{equation}
where 
\begin{equation}
    \mathcal{L}^+ = \sum\limits_j \frac{1}{\lambda_j} |x_j\rangle\langle y_j|
\end{equation}
is the Drazin inverse of the Liouvillian $\mathcal{L}$. 
Since $\mathcal{D}$ is traceless, when plugging Eq.~\eqref{eq: intergral final} into Eq.~\eqref{SM_deriv_partial1}, the term proportional to $\tau^2$ vanishes, and we obtain 
\begin{equation}\label{SM_deriv_final}
    \partial_\theta \text{E}_\theta [N(\tau)]_{\theta = 0} = \tau J - \tau \sum\limits_k \nu_k \Tr \{ \op{L}_k^\dagger \op{L}_k \mathcal{L}^+ \mathcal{D} \op{\rho}_0 \}. 
\end{equation}
Finally, using the facts that $\mathcal{L} = \mathcal{H} + \mathcal{D}$ and that $\op{\rho}_0$ is the steady state $\hat{\rho}_{\rm ss}$, we obtain
\begin{equation}
\label{eq:psider}
    \partial_\theta \text{E}_\theta [N(\tau)]_{\theta = 0} = \tau J \big(1 + \psi\big),
\end{equation}
where the factor $\psi$ is given by Eq.~\eqref{eq: psi} in the main text, reproduced here for convenience:
\begin{equation}\label{SM_psi_final}
    \psi = \frac{1}{J} \Tr \{ \mathcal{J} \mathcal{L}^+ \mathcal{H} \op{\rho}_{\rm ss} \}.
\end{equation}
Here we have employed the current superoperator, $\mathcal{J} \op{\rho} = \sum_k \nu_k L_k \op{\rho} L_k^\dagger$, see Eq.~\eqref{eq: Current op} in the main text.

For a diffusive current, Eq.~\eqref{eq: expectation theta} takes the form
\begin{equation}
    \text{E}_\theta[N(\tau)] = \sqrt{1+\theta} \int_0^\tau dt~\nu_k \Tr \{ \mc{J}_{\rm d} e^{\mathcal{L}_\theta t } \op{\rho}_0 \}.    
\end{equation}
One can notice that the only difference is the factor in front. As a result, Eq.~\eqref{SM_deriv_final} becomes  
\begin{equation}\label{SM_deriv_final_diff}
    \partial_\theta \text{E}_\theta [N(\tau)]_{\theta = 0} = \frac{1}{2} \tau J_{\rm d} - \tau \Tr \{ \mc{J}_{\rm d} \mathcal{L}^+ \mathcal{D} \op{\rho}_0 \}, 
\end{equation}
where we used $\partial_\theta \sqrt{1 + \theta}|_{\theta = 0} = 1/2$. The remaining calculations remain the same.

\section{II. Restricting $\psi$ for series of conductances}

Here we consider currents that can be written as a series of conductances
\begin{equation}
    \label{eq:currser}
    J = \left(\sum_j \gamma_j+\sum_k W_k\right)^{-1},
\end{equation}
where $\gamma_j$ are proportional to the rates occuring in the Lindblad jump operators, whereas 
\begin{equation}
\label{eq:pertrates}
    W_k = \frac{g_k^2}{\Gamma_k},
\end{equation}
are rates that follow from a perturbative treatment of coherent tunneling terms. In Eq.~\eqref{eq:pertrates}, $g_k$ is proportional to the coherent tunneling rates, while $\Gamma_k$ is proportional to the Lindblad jump rates. The deformation used to derive the $\psi$-KUR thus results in 
\begin{equation}
     \label{eq:defrates}
     \gamma_l\rightarrow \gamma_l(1+\theta),\hspace{2cm}W_k\rightarrow W_k/(1+\theta)\simeq W_k(1-\theta),
\end{equation}
where the last relation holds for small $\theta$. To linear order in $\theta$, the deformed integrated current follows from Eqs.~\eqref{eq:currser} and \eqref{eq:defrates} as
\begin{equation}
    E_\theta[N(\theta)] = J\tau\left(1+\theta\alpha\right),\hspace{2cm}\alpha = \frac{\sum_l \gamma_l^{-1}-\sum_k W_k^{-1}}{\sum_l \gamma_l^{-1}+\sum_k W_k^{-1}}.
\end{equation}
Since all the rates are positive, $-1\leq \alpha\leq 1$. From Eq.~\eqref{eq:psider}, we find $\alpha-1 = \psi$, implying that $-2\leq \psi\leq 0$.

\section{III. Quantum KUR of Ref.~\cite{Hasegawa_2020}}
\label{supp:Hasegawa}
The quantum correction $\chi$ in the quantum KUR in Eq.~\eqref{eq: Hasegawa KUR} is given by~\cite{Hasegawa_2020}
\begin{equation}
    \chi = -4 \left( \Tr\{\mc{K}_L \mc{L}^+\mc{K}_R \op{\rho}_{\rm ss} \} +  \Tr\{\mc{K}_R \mc{L}^+\mc{K}_L \op{\rho}_{\rm ss} \} \right),
\end{equation}
where
\begin{equation}
\begin{split}
    \mc{K}_L \op{\rho} =& -i\op{H} \op{\rho} + \frac{1}{2} \sum_k \left( \op{L}_k \op{\rho} \op{L}_k^\dagger - \op{L}_k^\dagger \op{L}_k \op{\rho}  \right), \\
    \mc{K}_R \op{\rho} =& i \op{\rho} \op{H}  + \frac{1}{2} \sum_k \left( \op{L}_k \op{\rho} \op{L}_k^\dagger -  \op{\rho} \op{L}_k^\dagger \op{L}_k  \right).
\end{split}
\end{equation}

\section{IV. Vectorization and Drazin inverse}
\label{supp:VecDraz}
\subsection{Vectorization}
In order to analytically compute fluctuations of currents and the KUR bounds, we express the density matrix of the system, $\op{\rho}$, as a matrix
\begin{equation}
    \op{\rho} = \sum_{i, j} \rho_{ij} \ket{i}\bra{j}
\end{equation}
in a basis $\{\ket{i}, \ket{j} \}$ of the Hilbert space
and write the Lindblad master equation~\eqref{eq: Lindblad} in the vectorized form~\cite{Landi_2023}:
\begin{equation}
    \frac{d}{dt} \ket{\rho} = \mc{L} \ket{\rho},
\end{equation}
where $\ket{\rho}$ is a vectorized density matrix.
The vectorization of the density matrix follows a convention of stacking the columns on top of each other such that the first one is on the top, i.e. $\ket{\rho} = [\rho_{11}, \rho_{21}, \rho_{31}, ..., \rho_{12}, \rho_{22}, \rho_{32}, ... ]^T$. The vectorized Liouvillian $\mc{L}$ can be obtained using the following relation for matrices $M_{1, 2}$:
\begin{equation}
    M_1 \op{\rho} M_2 \to M_2^T \otimes M_1 \ket{\rho}.   
\end{equation}
The steady state $\ket{\rho}$ is the eigenvector of $\mc{L}$ corresponding to the zero eigenvalue. The remaining superoperators that are necessary to compute $\psi$ such as $\mc{H}$ and $\mc{J}$ can be vectorized using the same technique.

\subsection{Drazin inverse}
The Drazin inverse $\mc{L}^+$~\cite{Mandal_2016} of the Liouvillian $\mc{L}$ can be found from the Moore-Penrose inverse $\mc{L}^{\rm MP}$ by projecting away the null space, i.e.,
\begin{equation}
    \mc{L}^+ = (\op{I} -\mc{P}) \mc{L}^{\rm MP} (\op{I} - \mc{P}),
\end{equation}
where $\op{I}$ is an identity operator and $\mc{P} \bullet := \op{\rho}_{\rm ss} \Tr \{ \bullet\} $. The Drazin inverse can be directly computed using the formula
\begin{equation}
    \mc{L}^+ = - \int_0^\infty ds e^{\mc{L}s} (\op{I} - \mc{P}).
\end{equation}
In the vectorized notation, $\mc{P}$ becomes $\ket{\rho} \bra{1}$, where $\ket{\rho}$ and $\bra{1}$ are right and left eigenvectors of $\mc{L}$ corresponding to the zero eigenvalue, respectively. The vectorized Drazin inverse $\mc{L}^+$ applied to a given $\ket{x}$ can be evaluated as $\mc{L}^+ \ket{x} = \left(I - \ket{\rho} \bra{1} \right) \ket{y}$, where $\ket{y}$ is a solution to $\mc{L} \ket{y} = \left(I - \ket{\rho} \bra{1} \right) \ket{x}$, and $I$ is a vectorized $\op{I}$.%~\cite{Flindt_2010, Landi_2023}.

\subsection{Noise $D$}

To compute the noise $D$, see Eq.~\eqref{eq: Current} in the main text, we may employ a method of full counting statistics~\cite{Schaller_2013}, which we briefly summarize here. We introduce a counting field $\chi$ and a generalized master equation
\begin{equation}
    \mc{L}_\chi \op{\rho}_t(\chi) \equiv \mc{L} \op{\rho}_t(\chi) + \sum_k (e^{i \nu_k \chi} -1) \op{L}_k  \op{\rho}_t(\chi) \op{L}_k^\dagger, 
\end{equation}
where $\mc{L}$ is the Liouvillian, see Eq.~\eqref{eq: Lindblad} in the main text. For the initial state $\op{\rho}_0$ that is assumed to be the steady state of  $\mc{L}$, a formal solution is given by $\op{\rho}_t(\chi) = e^{\mc{L}_\chi t} \op{\rho}_0$. %$\text{E}[N(t)]$ and $\text{Var}[N(t)]$ are given by, respectively, the first and second cumulant of corresponding to the cumulant generating function $\lambda(\chi) t + \beta(\chi)$, where $\lambda(\chi)$ is the eigenvalue of $\mc{L}_\chi$ with a largest real part and $\beta(\chi)$ is a polynomial. 
In the long time limit, the average current $J$ and the noise $D$ are given by
\begin{equation}
    J = -i\partial_\chi \lambda(\chi) |_{\chi = 0},\hspace{2cm}D = -\partial^2_\chi \lambda(\chi) |_{\chi = 0},
\end{equation}
where $\lambda(\chi)$ is the eigenvalue of $\mc{L}_\chi$ with a largest real part.

The noise $D$ for the quantum jump unravelling can be explicitly evaluated using the expression~\cite{Landi_2023}
\begin{equation}
\label{eq: D formula}
    D = \sum_k \Tr \{\nu_k^2 \op{L}_k \op{\rho} \op{L}_k^\dagger \} - 2\Tr \{ \mc{J} \mc{L}^+ \mc{J} \op{\rho} \}.
\end{equation}
For the diffusive unravelling, $D$ can be obtained using the formula~\cite{Landi_2023}
\begin{equation}
\label{eq: D formula diff}
    D = \sum_k \nu_k^2  - 2\Tr \{ \mc{J}_{\rm d} \mc{L}^+ \mc{J}_{\rm d} \op{\rho} \}.
\end{equation}

\section{V. Double quantum dot}
\label{supp:DQD}
\subsection{Time evolution and denisty matrix}
Fluctuations of the electron current passing through the DQD have been investigated recently~\cite{Prech_2023}. Using the basis $\{|n_L, n_R \rangle \equiv (\hat{c}_L^{\dagger})^{n_L} (\hat{c}_R^{\dagger})^{n_R} |0, 0 \rangle \}$, which corresponds to the occupation of each quantum dot, the steady-state density matrix has the following form:
\begin{equation}
\label{eq:Density Matrix}
    \hat{\rho} = 
    \begin{pmatrix}
        p_0 & 0 & 0 & 0 \\
        0 &  p_L &  \alpha & 0 \\
        0 &  \alpha^{*} &  p_R & 0 \\
        0 & 0 & 0 &  p_D
    \end{pmatrix}.
\end{equation}
The expression for the vectorized Liouvillian corresponding to Eq.~\eqref{eq: Liouvillian dqd} in the main text, written in the coefficient basis $\{p_0, p_L, p_R, p_D, \text{Re}[\alpha], \text{Im}[\alpha]\}$, reads
     \begin{equation}
\label{eq:Li Local}
\mc{L} =
\footnotesize
\begin{pmatrix}
        -f_L \gamma_L - f_R \gamma_R & (1-f_L)\gamma_L & (1-f_R)\gamma_R & 0 & 0 & 0 \\
         f_L \gamma_L & -(1-f_L)\gamma_L - f_R \gamma_R & 0 & (1-f_R)\gamma_R & 0 & -2g \\
        f_R \gamma_R & 0 & - f_L \gamma_L - (1-f_R) \gamma_R & (1-f_L)\gamma_L & 0 & 2g \\
        0 & f_R \gamma_R & f_L \gamma_L & -(1-f_L)\gamma_L - (1-f_R) \gamma_R & 0 & 0 \\
        0 & 0 & 0 & 0 & -\frac{\gamma_L + \gamma_R + 2 \Gamma}{2} & 0 \\
        0 & g & -g & 0 & 0 & -\frac{\gamma_L + \gamma_R + 2 \Gamma}{2}
    \end{pmatrix}
    \normalsize
    .
\end{equation}
The corresponding steady state is given by
%\begin{equation}
%\label{eq:Coefficients local}
%    \begin{split}
%        &p_0
%        = \frac{4g^2 (1-\bar{f})^2  + (1-f_L) (1-f_R) \gamma_L \gamma_R }{4g^2 +  \gamma_L \gamma_R},\\
%        &p_D
%        = \frac{4g^2 \bar{f}^2  + f_L f_R \gamma_L \gamma_R }{ 4g^2 +  \gamma_L \gamma_R},\\
%        &p_L
%        = \frac{4g^2 \bar{f}(1-\bar{f})  + f_L (1-f_R) \gamma_L \gamma_R}{4g^2 +  \gamma_L \gamma_R},\\
%        &p_R 
%        = \frac{4g^2 \bar{f}(1-\bar{f})  + (1-f_L) f_R \gamma_L \gamma_R }{4g^2 +  \gamma_L \gamma_R},\\
%        &\alpha
%        = \frac{2ig(f_L - f_R)\gamma_L \gamma_R}{4g^2 (\gamma_L + \gamma_R) +  \gamma_L \gamma_R(\gamma_L + \gamma_R + 2 \Gamma)},\\
%    \end{split}
%\end{equation}
\begin{equation}
\label{eq:Coefficients local}
    \begin{split}
        &p_0
        = \frac{4g^2 (1-\bar{f})^2  (\gamma_L + \gamma_R) + (1-f_L) (1-f_R) \gamma_L \gamma_R (\gamma_L + \gamma_R + 2 \Gamma) }{4g^2 (\gamma_L + \gamma_R) +  \gamma_L \gamma_R(\gamma_L + \gamma_R + 2 \Gamma)},\\
        &p_D
        = \frac{4g^2 \bar{f}^2 (\gamma_L + \gamma_R)  + f_L f_R \gamma_L \gamma_R (\gamma_L + \gamma_R + 2 \Gamma) }{4g^2 (\gamma_L + \gamma_R) +  \gamma_L \gamma_R(\gamma_L + \gamma_R + 2 \Gamma)},\\
        &p_L
        = \frac{4g^2 \bar{f}(1-\bar{f}) (\gamma_L + \gamma_R)  + f_L (1-f_R) \gamma_L \gamma_R (\gamma_L + \gamma_R + 2 \Gamma) }{4g^2 (\gamma_L + \gamma_R) +  \gamma_L \gamma_R(\gamma_L + \gamma_R + 2 \Gamma)},\\
        &p_R 
        = \frac{4g^2 \bar{f}(1-\bar{f}) (\gamma_L + \gamma_R)  + (1-f_L) f_R \gamma_L \gamma_R (\gamma_L + \gamma_R + 2 \Gamma) }{4g^2 (\gamma_L + \gamma_R) +  \gamma_L \gamma_R(\gamma_L + \gamma_R + 2 \Gamma)},\\
        &\alpha
        = \frac{2ig(f_L - f_R)\gamma_L \gamma_R}{4g^2 (\gamma_L + \gamma_R) +  \gamma_L \gamma_R(\gamma_L + \gamma_R + 2 \Gamma)},\\
    \end{split}
\end{equation}
where $\bar{f} \equiv \frac{f_L \gamma_L + f_R \gamma_R}{\gamma_L + \gamma_R}$.
The $l_1$ norm of coherence~\cite{Baumgratz_2014} in the occupation basis is defined as $   \mc{C} = |\alpha|$.%, which for the DQD is provided in Eq.~\eqref{eq: norm dqd}, is defined as
 
%with $\alpha$.% given in Eq.~\eqref{eq:Coefficients local}.

\subsection{Jump unravelling}
We consider a current passing through the DQD, which corresponds to, for instance, setting the weight associated with $\op{c}_L^\dagger$ to $\nu = 1$, the weight associated with  $\op{c}_L$ to $\nu = -1$, and all remaining weights to $\nu = 0$. The average value of the current $J = \gamma_L \Tr \{ f_L \op{c}_L^\dagger \op{\rho} \op{c}_L - (1-f_L) \op{c}_L \op{\rho} \op{c}_L^\dagger \}$ is given by
\begin{equation}
\label{eq: Current local}
    J = \frac{4g^2 (f_L - f_R) \gamma_L \gamma_R}{4g^2 (\gamma_L + \gamma_R)+ \gamma_L \gamma_R (\gamma_L + \gamma_R + 2 \Gamma)}.
\end{equation}
The dynamical activity reads
\begin{equation}
\label{eq: M dqd}
    A = \frac{8g^2 (\gamma_L + \gamma_R)^2\bar{f}(1-\bar{f} + \Gamma/(\gamma_L + \gamma_R)) +  \gamma_L \gamma_R (\gamma_L + \gamma_R + 2 \Gamma) \left(2\gamma_L f_L(1-f_L) + 2\gamma_R f_R(1-f_R) + \Gamma (f_L + f_R)\right)}{4g^2 (\gamma_L + \gamma_R) + \gamma_L \gamma_R (\gamma_L + \gamma_R + 2 \Gamma)}.
\end{equation}
The analytical expression obtained for the noise with Eq.~\eqref{eq: D formula} is given by
\begin{equation}
\label{eq:Variance local e}
    \begin{split}
        D = &\frac{4g^2 (f_L + f_R - 2f_L f_R) \gamma_L \gamma_R}{4g^2 (\gamma_L + \gamma_R) + \gamma_L \gamma_R (\gamma_L + \gamma_R + 2 \Gamma)} + J \frac{32 g^2 (f_L - f_R ) \gamma_L \gamma_R(\gamma_L + \gamma_R + \Gamma)}{(\gamma_L + \gamma_R)(4g^2 (\gamma_L + \gamma_R) + \gamma_L \gamma_R (\gamma_L + \gamma_R + 2 \Gamma))} \\
        &-2 J^2 \frac{4 g^2 (\gamma_L + \gamma_R )(5\gamma_L + 5\gamma_R + 6\Gamma) + 2(\gamma_L + \gamma_R + 2 \Gamma)(2 \Gamma(\gamma_L^2 + 3 \gamma_L \gamma_R + \gamma_R^2) + (\gamma_L + \gamma_R ) (\gamma_L^2 + 7 \gamma_L \gamma_R + \gamma_R^2) )}{(\gamma_L + \gamma_R)(4g^2 (\gamma_L + \gamma_R) + \gamma_L \gamma_R (\gamma_L + \gamma_R + 2 \Gamma))}.
        \end{split}
\end{equation}
%where $J$ is given in Eq.~\eqref{eq: Current local}.

%The quantum correction $\psi$~\eqref{eq: psi} is given by a compact formula:
%\begin{equation}
%    \psi = \frac{4g^2 - \gamma_L \gamma_R}{4g^2 + \gamma_L \gamma_R} -1,
%\end{equation}
%which leads to
%\begin{equation}
%    (1+\psi)^2 = \frac{(4g^2 - \gamma_L \gamma_R)^2}{(4g^2 + \gamma_L \gamma_R)^2}.
%\end{equation}

\subsection{Diffusive unravelling}
For the diffusive unravelling, the average current $J_{\rm d} =  \sqrt{2\Gamma} \Tr\{ (\op{c}_L^\dagger \op{c}_L - \op{c}_R^\dagger \op{c}_R) \op{\rho} \} $ is given by
\begin{equation}
    J_{\rm d} = \frac{ \sqrt{2\Gamma} (f_L - f_R) \gamma_L \gamma_R (\gamma_L + \gamma_R + 2 \Gamma)}{4g^2 (\gamma_L + \gamma_R) + \gamma_L \gamma_R (\gamma_L + \gamma_R + 2 \Gamma)},
\end{equation}
and the dynamical activity is still given in Eq.~\eqref{eq: M dqd}. The expression for the noise $D$, which is computed using Eq.~\eqref{eq: D formula diff} and Eq.~$\eqref{eq: Current diff}$ in the main text, with $\nu = 1$, $\phi = 0$, and $\op{L} = \op{c}_L^\dagger \op{c}_L - \op{c}_R^\dagger \op{c}_R$, is omitted here due to its length.
%of the expression, here we refrain from writing the entire analytical expressions explicitly.
%The expression for the factor $\psi$~\eqref{eq: psi dqd diff} reads
%\begin{equation}
%    \psi = \frac{8g^2 (\gamma_L + \gamma_R)}{4g^2 (\gamma_L + \gamma_R) + \gamma_L \gamma_R (\gamma_l + \gamma_R + 2 \Gamma)}.
%\end{equation}

\subsection{Factor $\psi$ for unidirectional counting}
In the case of the jump unravelling, if we count only electrons entering the system from the left reservoir, which corresponds to the current $J = \gamma_L f_L \text{Tr}\{\op{c}_L^\dagger \op{\rho} \op{c}_L \}$, we find the factor $\psi$, without dephasing for simplicity $(\Gamma = 0)$:
\begin{equation}
    \psi = \frac{8g^2 (f_L - f_R) \gamma_L \gamma_R^2}{(4g^2 + \gamma_L \gamma_R) ( (f_L - 1)\gamma_L \gamma_R (\gamma_l + \gamma_R) + 4g^2 ((f_L - 1)\gamma_L + (f_R - 1)\gamma_R ) )} ,
\end{equation}
which can be positive.

\section{VI. Effective classical model of the double quantum dot}
\label{supp: Clas dqd}
\subsection{Rate equation}
The time evolution of the vector of populations $\vec{p} = [p_{0}, p_{L}, p_R, p_D]^T$ in the classical model given in Eq.~\eqref{eq: Rate equation} in the main text is described by the transition matrix
\begin{equation}
\label{eq:W}
W =
\footnotesize
\begin{pmatrix}
        -f_L \gamma_L - f_R \gamma_R & (1-f_L)\gamma_L & (1-f_R)\gamma_R & 0 \\
        f_L \gamma_L & -(1-f_L)\gamma_L - f_R \gamma_R- W_{RL} & W_{LR} & (1-f_R)\gamma_R \\
        f_R \gamma_R & W_{RL} & - f_L \gamma_L - (1-f_R)\gamma_R - W_{LR} & (1-f_L)\gamma_L  \\
        0 & f_R \gamma_R & f_L \gamma_L & -(1-f_L)\gamma_L - (1-f_R) \gamma_R  
    \end{pmatrix}
    \normalsize
    ,
\end{equation}
where $W_{LR} = W_{RL}= \frac{4g^2}{\gamma_L + \gamma_R +2 \Gamma}$, which has already been introduced in Eq.~\eqref{eq: WLR} in the main text, replaces the coherent tunneling. The remaining elements of the matrix $W$, which correspond to transitions due to the reservoirs, are obtained directly from the dissipative part of the Lindblad master equation~\eqref{eq: Liouvillian dqd}. The effective classical model captures both the average current $J$~\eqref{eq: Current local} across the DQD and the steady state of the quantum model, meaning that the steady state of $W$ agrees with the diagonal elements of the steady-state density matrix~\eqref{eq:Coefficients local}.

The expressions for the noise in the classical model, which is different from the quantum expression~\eqref{eq:Variance local e}, can be found using Eq.~\eqref{eq: D formula} adapted to the classical rate equation:
\begin{equation}
    \label{Variance dqd classical}
    \begin{split}
        D = &\frac{4g^2 (f_L + f_R - 2f_L f_R) \gamma_L \gamma_R}{4g^2 (\gamma_L + \gamma_R) + \gamma_L \gamma_R (\gamma_L + \gamma_R + 2 \Gamma)} + J \frac{16 g^2 (f_L - f_R ) \gamma_L \gamma_R}{4g^2 (\gamma_L + \gamma_R) + \gamma_L \gamma_R (\gamma_L + \gamma_R + 2 \Gamma)} \\
        &-2 J^2 \frac{12 g^2 (\gamma_L + \gamma_R ) + (\gamma_L + \gamma_R + 2 \Gamma)(\gamma_L^2 + 3 \gamma_L \gamma_R + \gamma_R^2)}{(\gamma_L + \gamma_R)(4g^2 (\gamma_L + \gamma_R) + \gamma_L \gamma_R (\gamma_L + \gamma_R + 2 \Gamma))},
        \end{split}
\end{equation}
where $J$ is given in Eq.~\eqref{eq: Current local}.
The dynamical activity of the classical model, defined as $A_{\rm cl} = \sum_{j \neq k} W_{kj}p_j$, is given by the expression
\begin{equation}
    A_{\rm cl} = A + \left( \frac{4g^2}{\gamma_L + \gamma_R + 2 \Gamma} -\frac{\Gamma}{2} \right) (p_L + p_R) ,
\end{equation}
where $p_{L, R}$ are provided in Eq.~\eqref{eq:Coefficients local}, and $A$ is the dynamical activity of the quantum model~\eqref{eq: M dqd}. The term proportional to $\Gamma$ is subtracted, because dephaing jumps contribute to the dynamical activity in the quantum model, but they do not in th classical rate equation. Hence, for small $g$ as well as $\Gamma$, we have $A \simeq A_{\rm cl}$.

\subsection{Adiabatic elimination for large $g$}
Increasing the value of $g$ gives rise to a larger dynamical activity $A_{\rm cl}$ due to frequent transitions of electrons between the left and right quantum dots. For large $g$, the rate of the interdot transitions is much higher than the rate of the transitions in and out of the system, leading to $p_L \simeq p_R$. However, these fast interdot oscillations do not influence fluctuations of the current and are, therefore, redundant in the description of the dynamics of the system. The effective dynamics is described by the three-state model, where we coarse-grain the states with a single electron occupation, resulting in the rate equation for the vector $\vec{p} = [p_0, p_L + p_R, p_D]$. The associated transition matrix is obtained using the adibatic elimination as follows. First, we make a basis change $\vec{p} = [p_0, p_L, p_R, p_D] \to  \vec{p} = [p_0, p_L + p_R, p_L - p_R, p_D]$. The corresponding transition matrix reads
\begin{equation}
\tilde{W} =
\footnotesize
\begin{pmatrix}
        -f_L \gamma_L - f_R \gamma_R & \frac{(1-f_L)\gamma_L + (1-f_R)\gamma_R}{2} & \frac{(1-f_L)\gamma_L - (1-f_R)\gamma_R}{2} & 0 \\
        f_L \gamma_L + f_R \gamma_R & -\frac{\gamma_L + \gamma_R}{2} & \frac{-\gamma_L (1-2f_L) + \gamma_R(1-2f_R)}{2} & (1-f_L)\gamma_L + (1-f_R)\gamma_R \\
        f_L \gamma_L - f_R \gamma_R &  \frac{-\gamma_L (1-2f_L) + \gamma_R(1-2f_R)}{2} & -\frac{\gamma_L + \gamma_R + 4 W_{LR}}{2} & -(1-f_L)\gamma_L + (1-f_R)\gamma_R  \\
        0 & \frac{f_L \gamma_L + f_R \gamma_R}{2} & \frac{-f_L \gamma_L + f_R \gamma_R}{2} & -(1-f_L)\gamma_L - (1-f_R) \gamma_R  
    \end{pmatrix}
    \normalsize
    .
\end{equation}
Let $+$ and $-$ denote the states corresponding to $p_L + p_R $ and $p_L - p_R$, respectively. In the steady state, the expression for $p_-$ reads $p_- = -(\tilde{W}_{-0}p_0 + \tilde{W}_{-+}p_+ + \tilde{W}_{-D}p_D )/\tilde{W}_{--}$. We may insert it into each equation $\frac{d}{dt} p_k = \sum_{j \in \{0, +, D \}} \tilde{W}_{kj} p_j + \tilde{W}_{k-} p_-$ describing the time evolution of the probabilities $p_k$ for $k \in \{0, +, D \}$, which results in a coupled system of differential equations for $p_{0, +, D}$. It can be expressed as a new rate equation $\frac{d}{dt} \vec{p} = \mathbb{W} \vec{p}$, with $\vec{p} = [p_0, p_+, p_D]$, where
\begin{equation} \label{eq: W adiab}
     \mathbb{W}_{kj} = \tilde{W}_{kj} - \frac{ \tilde{W}_{k-} \tilde{W}_{-j} }{\tilde{W}_{--}}.
\end{equation}
The corresponding dynamical activity is given by $A_{\rm ad} = \sum_{k, j \in \{0, +, D \}: k\neq j} \mathbb{W}_{kj}p_j$, resulting in
\begin{equation}
    A_{\rm ad} = (\gamma_L + \gamma_R)\frac{128g^4 \bar{f} (1- \bar{f}) +g^2 \mathbb{P}_2 + \mathbb{P}_4}{(4g^2 + \gamma_L \gamma_R)(16g^2 +(\gamma_L +\gamma_R)^2)},
\end{equation}
where $\mathbb{P}_{2, 4}$ are two different polynomials in $\gamma_L$ and $\gamma_R$ (with products of $f_L$ and $f_R$ being coefficients) of the order $2$ and $4$, respectively. Therefore, for large $g/\gamma_{L, R}$, $A_{\rm ad}$ reduces to $A_{\rm ad} \simeq 2 (\gamma_L + \gamma_R) \bar{f} (1- \bar{f})$, which
tends to $A$ in the same limit and small $\Gamma$, cf. Eq.~\eqref{eq: M dqd}.% is in agreement with the dynamical activity $A$~\eqref{eq: M dqd} of the Lindblad master equation, which is illustrated in Fig.~\ref{fig:ActAdiabatic} for the same set of parameters as in Fig.~\ref{fig:panel1}(a).

%\begin{figure*}[h]%{L}{0.4\textwidth}
%    \centering
%    \includegraphics[width=0.4\textwidth]{AvsAad.png}
%    \caption{Comparison of the dynamical activity $A$~\eqref{eq: M} of the Lindblad master equation given in Eq.~\eqref{eq: M dqd} (solid line) and the dynamical activity $A_{\rm cl}$ of the three state model~\eqref{eq: W adiab} obtained using the adiabatic elimination (dashed line) as a function of $g/\gamma$ with $\gamma = \gamma_L = \gamma_R$. Parameters correspond to Fig.~\ref{fig:panel1}(a): $\beta_L \mu_L = - \beta_R \mu_R = 7$, $\epsilon = 0$, and $\Gamma = 0$.}
%    \label{fig:ActAdiabatic}
%\end{figure*}

\section{VII. Random network of states}
\label{supp:network}

We consider a 5-level quantum system in the basis $\{ \ket{n} \}$. The Hamiltonian of the system is given by
\begin{equation}
    \op{H} = \sum_{n\neq k} g_{nk} \ket{n} \bra{k}.
\end{equation}
Each transition $n \leftrightarrow k$ is coherently connected with $50\%$ probability. For each connected transition, values of $g_{nk} = g_{kn}$ are sampled form a uniform distribution $U[0, 3]$ with a probability density
\begin{equation}
    p_{U[a, b]}(x) =  \begin{cases}
        &\frac{1}{b-a} \text{ 
  for   } a \leq x \leq b,\\
  & 0 \text{   otherwise}. 
    \end{cases}
\end{equation}
In the case of no coherent transition, we set $g_{nk} = 0$, and this connection is mediated by a reservoir-like transition, giving rise to the 
$ \gamma_{nk}\mc{D}[ \ket{n} \bra{k}] +  \gamma_{kn} \mc{D}[ \ket{k} \bra{n}]$ contribution in the Lindblad master equation of the system. Each $\gamma_{nk}$ is sampled from $U[0, 3]$, whereas  $\gamma_{kn} = \gamma_{nk} e^{-\sigma_{nk}}$, where $\sigma_{nk}$ is sampled from $U[3, 5]$ to ensure a strong bias.
Current is calculated along a single reservoir-like transition with antisymmetric weights, $J = \gamma_{10} \langle 0 | \hat{\rho}_{\rm ss} | 0 \rangle - \gamma_{01} \langle 1 | \hat{\rho}_{\rm ss} | 1 \rangle$.
We simulated $1000$ random network configurations and computed $J, D, A$, $\psi$, and $\chi$ for each of them. The scatter plot of $DA/J^2$ against $(1 + \psi)^2$ (red circles) and against $A/(A + \chi)$ (green circles) is shown in Fig.~\ref{fig: random network} in the main text.

\section{VIII. Further example: coherently driven qubit}
\label{supp: coh dr qub}
To further illustrate our results, we investigate a coherently driven qubit with Hamiltonian 
\begin{equation}
\label{eq: H Qubit}
    \op{H} = \frac{\Delta}{2} \op{\sigma}_z + \Omega \op{\sigma}_x,
\end{equation}
where $\Delta$ is the detuning, $\Omega$ the Rabi frequency, and $\op{\sigma}_{x, y, z}$ are Pauli matrices. The qubit is weakly coupled to a thermal reservoir, and its time evolution in the rotating frame may be described by the Liouvillian
\begin{equation}
\label{eq: Liouvillian Qubit}
    \partial_t\op{\rho}=\mathcal{L} \op{\rho} = -i[\op{H}, \op{\rho}] + \kappa \bar{n} D[\op{\sigma}_+] \op{\rho} + \kappa (\bar{n}+1)D[\op{\sigma}_-] \op{\rho} 
\end{equation}
where $\bar{n} \equiv (\exp{(\beta \omega)} - 1)^{-1}$ denotes the Bose-Einstein occupation, $\beta$ is the inverse temperature of the reservoir, $\kappa$ is the strength of the coupling to the reservoir, and $\omega$ is the energy gap of the qubit in the Sch\"{o}dinger picture.

The Hamiltonian~\eqref{eq: H Qubit} may implemented with a two-level atom characterized by a transition frequency $\omega$ and driven by a coherent laser field with a frequency $\omega_d$. The corresponding time-dependent Hamiltonian $\op{H}(t) = \op{H}_0 + \op{V}(t)$ consists of a bare part, $\op{H}_0 = \omega \op{\sigma}_+ \op{\sigma}_-$, and a driving contribution, $ \op{V}(t) = \Omega (\op{\sigma}_+ \exp{(-i\omega_d t)}+ \op{\sigma}_- \exp{(i\omega_d t)})$, where $\op{\sigma}_\pm = (\op{\sigma}_x \pm \op{\sigma}_y)/2$ is a raising (lowering) operator. Upon moving to the rotating frame with respect to $\exp{(i \omega_d t \op{\sigma}_+ \op{\sigma}_-)}$, we obtain the Hamiltonian in Eq.~\eqref{eq: H Qubit}, where the detuning is given by $\Delta = \omega - \omega_d$.

In the basis $\{ \ket{0} , \ket{1} \}$, which corresponds to, respectively, the ground and the excited state, the steady-state density matrix can be written as
\begin{equation}
    \hat{\rho} = 
    \begin{pmatrix}
        \rho_{0 0} & \rho_{0 1} \\
        \rho_{1 0} &  \rho_{1 1} 
    \end{pmatrix}.
\end{equation}
The vectorized  Liouvillian~\eqref{eq: Liouvillian Qubit} expressed in the coefficient basis $\{\rho_{0 0}, \rho_{1 0}, \rho_{0 1}, \rho_{1 1} \}$ is given by
\begin{equation}
    \mc{L} = 
    \footnotesize
    \begin{pmatrix}
        -\kappa \bar{n} & -i \Omega & i \Omega & \kappa (\bar{n}+1) \\
        -i \Omega &  -i \Delta - \kappa (2\bar{n} +1)/2 &  0 & i \Omega \\
        i \Omega &  0 &  i \Delta - \kappa (2\bar{n} +1)/2 & -i \Omega \\
        \kappa \bar{n} & i \Omega & -i \Omega & -\kappa (\bar{n}+1)
    \end{pmatrix}.
    \normalsize
\end{equation}
The corresponding steady-state solution reads
%\begin{equation}
%    \begin{split}
%         \rho_{0 0} &= \frac{(1+\bar{n}) \left(\kappa^2 (1 + 2\bar{n})^2+4 \Delta ^2\right)+4 (1+2 \bar{n}) \Omega ^2}{(1+2 \bar{n}) \left(\kappa^2 (1 + 2\bar{n})^2+4 \left(\Delta ^2+2 \Omega ^2\right)\right)},  \\
%         \rho_{1 0} &= \frac{2 (-i \kappa(1+2\bar{n}) -2 \Delta ) \Omega }{(1+2 \bar{n}) \left(\kappa^2 (1 + 2\bar{n})^2+4 \left(\Delta ^2+2 \Omega ^2\right)\right)}, \\
%        \rho_{0 1} &= \frac{2  (i \kappa(1+2 \bar{n}) - 2 \Delta ) \Omega }{(1+2 \bar{n}) \left(\kappa^2 (1 + 2\bar{n})^2+4 \left(\Delta ^2+2 \Omega ^2\right)\right)},\\
%        \rho_{1 1} &= \frac{\bar{n} \left(\kappa^2 (1 + 2\bar{n})^2+4 \Delta ^2\right)+4 (1+2 \bar{n}) \Omega ^2}{(1+2 \bar{n}) \left(\kappa^2(1+2 \bar{n})^2+4 \left(\Delta ^2+2 \Omega ^2\right)\right)} , \\
%    \end{split}
%\end{equation}
\begin{equation}
\label{eq: steady state qubit}
    \begin{split}
         \rho_{0 0} &= \frac{(1+\bar{n}) \left(\mc{Q} - 8\Omega^2 \right)+4 (1+2 \bar{n}) \Omega ^2}{(1+2 \bar{n}) \mc{Q}},  \\
         \rho_{1 0} &= \frac{2 (-i \kappa(1+2\bar{n}) -2 \Delta ) \Omega }{(1+2 \bar{n}) \mc{Q}}, \\
        \rho_{0 1} &= \frac{2  (i \kappa(1+2 \bar{n}) - 2 \Delta ) \Omega }{(1+2 \bar{n}) \mc{Q}},\\
        \rho_{1 1} &= \frac{\bar{n} \left(\mc{Q} - 8\Omega^2 \right)+4 (1+2 \bar{n}) \Omega ^2}{(1+2 \bar{n}) \mc{Q} } , \\
    \end{split}
\end{equation}
where $\mc{Q} = \kappa^2(1+2 \bar{n})^2+4 \left(\Delta ^2+2 \Omega ^2\right)$.

\subsection{Jump current}
We consider the average current passing from the qubit to the reservoir, which corresponds to setting $\nu = 1$ for $\op{\sigma}_-$ and $\nu = -1$ for $\op{\sigma}_+$, i.e.,
\begin{equation}
    \mc{J} \op{\rho}_{\rm ss} = \kappa (\bar{n}+1) \op{\sigma}_- \op{\rho}_{\rm ss} \op{\sigma}_-^\dagger - \kappa \bar{n}  \op{\sigma}_+ \op{\rho}_{\rm ss} \op{\sigma}_+^\dagger .
\end{equation}
The analytical expression is given by
\begin{equation}
\label{eq: J qubit}
    J = \frac{4 \kappa \Omega^2}{\mc{Q}}.
\end{equation}
The dynamical activity of the qubit is given by the compact expression $A = \kappa \bar{n} \rho_{0 0} + \kappa(\bar{n} +1 ) \rho_{1 1}$. Inserting the steady-state solution~\eqref{eq: steady state qubit} leads to
\begin{equation}
\label{eq: M qubit}
    A = \frac{2 \kappa \bar{n}(1+\bar{n}) \left(\mc{Q} - 8 \Omega^2 \right)+4 \kappa (1+2 \bar{n})^2 \Omega ^2}{(1+2 \bar{n}) \mc{Q} }.
\end{equation}
The noise $D$ evaluated with Eq.~\eqref{eq: D formula} is given by 
\begin{equation}
\begin{split}
    D &= A +\\
    & - \frac{2 \kappa \left(\bar{n} (1+\bar{n}) \left( \mc{Q} - 8 \Omega^2 \right)^3+16 \bar{n} (1+\bar{n}) \left( \mc{Q} - 8 \Omega^2 \right)^2 \Omega ^2+16 \left( \kappa^2 (1+2 \bar{n})^2 (3+4 \bar{n} (1+\bar{n}))+4 (-1+4 \bar{n} (1+\bar{n})) \Delta ^2\right) \Omega ^4\right)}{(1 + 2 \bar{n}) \mc{Q}^3}.
\end{split}
\end{equation}
The expression for the factor $\psi$~\eqref{eq: psi} reads 
\begin{equation} \label{eq: psi qubit}
    \psi = \frac{-2 \kappa^2 (1 + 2 \bar{n})^2}{\mc{Q}}.
\end{equation}
%resulting in
%\begin{equation}
%    (1 + \psi)^2 = \frac{\bigg(4 \left(\Delta ^2+2 \Omega ^2\right)  - \kappa^2(1+2 \bar{n})^2 \bigg)^2}{\mc{Q}^2}.
%\end{equation}
%As far as the coefficient $\chi$ appearing in the Hasegawa's KUR~\eqref{eq: Hasegawa KUR} is concerned, here we omit writing the explicit formula.

\begin{figure*}[t]
    \centering
    \includegraphics[width=0.98\textwidth]{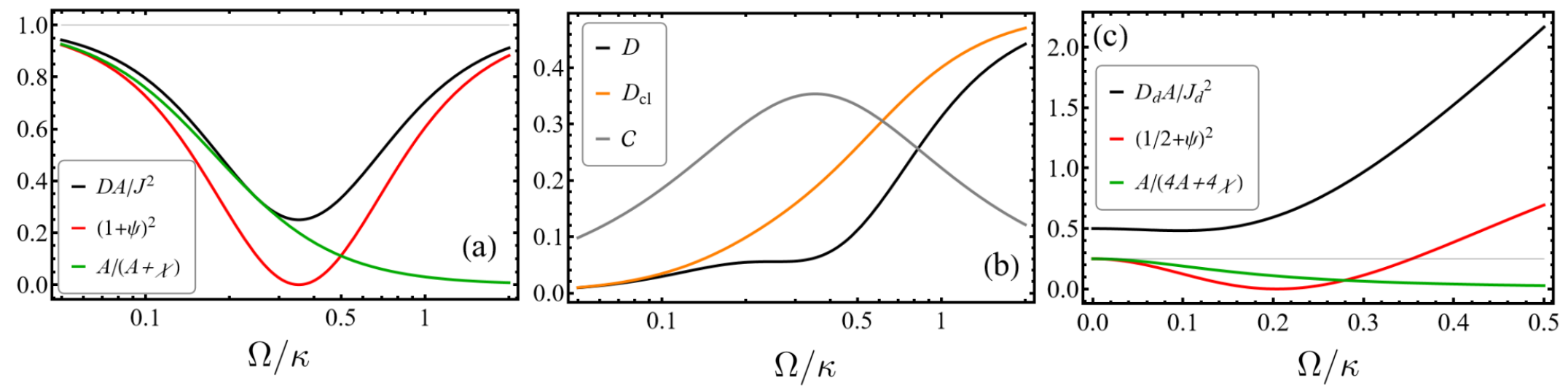}
    \caption{Quantum KUR in the coherently driven qubit. (a) Quantum jump unravelling: $DA/J^2$ (black), $(1 + \psi)^2$ bound of the $\psi$-KUR~\eqref{eq: Our KUR} (red) obtained with Eq.~\eqref{eq: psi qubit}, and $A/(A+\chi)$ bound of the KUR~\eqref{eq: Hasegawa KUR}.% as a function of $\Omega/\kappa$. The gray line is not to be numerically compared with the other lines.
    Parameters: $\bar{n} =0$ and $\Delta = 0$.
    (b) The noise $D$ for quantum model~\eqref{eq: Liouvillian Qubit} (black) and the classical model~\eqref{eq: rate qubit} (orange), and $l_1$ norm of coherence $\mc{C}$ with parameters of (a).  
    (c) Diffusive unravelling: analogous plots as in (a) for the diffusive measurement of the current in Eq.~\eqref{eq: curr diff qubit}. Parameters are the same as in (a).}
    \label{fig:panel2}
\end{figure*}

The $\psi$-KUR~\eqref{eq: Our KUR} is illustrated in Fig.~\ref{fig:panel2}(a), where we show $(1+\psi)^2$ (red line) bounding the ratio $DA/J^2$ (black line) from below and compare with the $A/(A + \chi)$ (green line). The observation made in the main text for the DQD that the dips in $(1+\psi)^2$ and $DA/J^2$ coincide carries over to the coherently driven qubit. We also show the $l_1$ norm of coherence, $\mc{C} = (|\rho_{01}| + |\rho_{10}|)/2$, in Fig.~\ref{fig:panel2}(b) as the gray line, whose peak appears in the same regime as the forementioned dips.

\subsection{Classical model}
Similarly to the DQD, we may consider the effective classical model governing the time-evolution of the populations $\vec{p} = [\rho_{00}, \rho_{11}]^T$ with a rate equation $\frac{d}{dt} \vec{p} = W \vec{p}$. The transition matrix for the qubit system is given by
\begin{equation} \label{eq: rate qubit}
    W = 
    \begin{pmatrix}
        -\kappa \bar{n} -  \Gamma_{\rm c} & \kappa (\bar{n}+1) + \Gamma_{\rm c} \\
        \kappa \bar{n} + \Gamma_{\rm c} & -\kappa (\bar{n}+1) - \Gamma_{\rm c}
    \end{pmatrix},
\end{equation}
where $\Gamma_{\rm c} = \frac{4\kappa (1 + 2 \bar{n}) \Omega^2}{\kappa^2 (1 + 2 \bar{n})^2 +4 \Delta^2}$. 
The steady state of the transition matrix $W$ and the average current $J$ are the same as the steady-state populations of the quantum model~\eqref{eq: steady state qubit} and the current in Eq.~\eqref{eq: J qubit}, respectively. The noise $D$, however, is different than in the quantum model: 
\begin{equation}
    D = \frac{4 \kappa \Omega ^2 \left(\left(\kappa^2 (1+2 \bar{n})^3+4 (1+2 \bar{n}) \Delta ^2\right)^2+8 (1+8 \bar{n} (1+\bar{n})) \left(\kappa^2 (1+2\bar{n})^2+4 \Delta ^2\right) \Omega
^2+64 (1+2 \bar{n})^2 \Omega ^4\right)}{(1+2 \bar{n}) \mc{Q}^3}.
\end{equation}
%The dynamical activity $A$ contains a contribution due to the classical rate $\Gamma_{\rm c}$ in addition to the interaction with the reservoir, which is encompassed by $M$ in Eq.~\eqref{eq: M qubit}:
%\begin{equation}
%    A = M + \Gamma_{\rm c} .
%\end{equation}

Like in the DQD case, there is a regime where $D$ of the classical rate equation fails to capture $D$ of the quantum model, which is illustrated in Fig.~\ref{fig:panel2}(b). This regime coincides with the regime where $DA/J^2$ and $(1+\psi)^2$ display dips in Fig.~\ref{fig:panel2}(a). This implies that likewise, for the coherently driven qubit, $\psi$ indicates a contributing role of coherence to overcoming the classical bound when the effective classical model is insufficient to describe the noise faithfully. Away from this regime, where the classical model provides a faithful description of $D$, the factor $(1 + \psi)^2$ approaches $1$.

\subsection{Diffusive current}

For the diffusive current we consider $\phi = \pi/2$ in Eq.~\eqref{eq: Current diff} in the main text, resulting in
\begin{equation} \label{eq: curr diff qubit}
    \mc{J}_{\rm d} \op{\rho}_{\rm ss} = \kappa (\bar{n}+1) \left( -i\op{\sigma}_- \op{\rho}_{\rm ss} +i \op{\rho}_{\rm ss} \op{\sigma}_-^\dagger \right) - \kappa \bar{n}  \left( -i \op{\sigma}_+ \op{\rho}_{\rm ss} +i \op{\rho}_{\rm ss} \op{\sigma}_+^\dagger \right) .
\end{equation}
The expression for the current reads
\begin{equation}
    J_{\rm d} = 4\kappa \Omega \frac{ \sqrt{\kappa \bar{n} } + \sqrt{\kappa (1 + \bar{n}) } }{\kappa^2 (1 + 2\bar{n})^2 + 4(\Delta^2 + 2 \Omega^2)}.
\end{equation}
We omit an explicit analytical expression for $D$ due to its complexity. We find
\begin{equation}
    \psi = \frac{\mc{Q} - 2 \kappa^2 (1 + 2 \bar{n})^2 }{\mc{Q}}.
\end{equation}
The KUR for the diffusive current is illustrated in Fig.~\ref{fig:panel2}(c), and the same conclusions as in the case of the DQD can be drawn.

\end{document}